\documentclass[aps,prb,twocolumn,superscriptaddress,showpacs,a4paper]{revtex4}
\usepackage{graphicx}
\usepackage{amssymb}
\usepackage{amsmath}
\usepackage{ulem}
\usepackage{color}
\usepackage{subfigure}
\usepackage{subeqnarray}
\usepackage[pdftex,breaklinks=true,colorlinks=true]{hyperref}
\usepackage[active]{srcltx}
\usepackage{multirow}

\newcommand{\qzero}{$\mathbf q=0$ }
\newcommand{\cuboc}{\textit{cuboc1} }
\newcommand{\cubocd}{\textit{cuboc2} }
\newcommand{\sqrtsqrt}{$\sqrt3\times \sqrt3$ }

\usepackage[utf8]{inputenc}

\begin{document}

\title{Time-reversal-symmetry-breaking chiral spin liquids: a projective symmetry group approach of bosonic mean-field theories}

\author{Laura Messio}
\affiliation{Institut de Physique Th\'eorique, CNRS, URA 2306, CEA, IPhT, 91191 Gif-sur-Yvette, France}

\author{Claire Lhuillier}
\affiliation{Laboratoire de Physique Th\'eorique de la Mati\`ere Condens\'ee, UMR 7600 CNRS, Universit\'e Pierre et Marie Curie, Paris VI, 75252 Paris Cedex 05, France}

\author{Gr\'egoire Misguich}
\affiliation{Institut de Physique Th\'eorique, CNRS, URA 2306, CEA, IPhT, 91191 Gif-sur-Yvette, France}

\begin{abstract}
Projective symmetry groups (PSG)
are the mathematical tools which allow to list and classify mean-field spin liquids (SL's) based on a parton construction.
The seminal work of Wen and its subsequent extension to bosons by Wang and Vishwanath
concerned the so-called \textit{symmetric} SL's: i.e. states  that break neither lattice symmetries nor time reversal invariance.
Here we generalize this approach to \textit{chiral} (time reversal symmetry breaking) SL's described in a Schwinger boson mean-field approach.
A special emphasis is put on frustrated lattices (triangular and kagome lattices), where the possibility of a chiral SL ground state has recently been discussed.
The PSG approach is detailed for the triangular lattice case. Results for other lattices are given in the appendices. The physical significance of gauge invariant
quantities called fluxes is discussed both in the classical limit and in the quantum SL and their expressions in terms of spin observables are given.

\end{abstract}

\maketitle

\section{Introduction}
Symmetry breaking is an ubiquitous feature of the low temperature behavior in condensed matter physics. Solids or N\'eel antiferromagnets are phases that break some essential symmetries of the physical laws:
translational symmetry or rotational spin symmetry. Understanding the nature of the broken symmetries, discrete or continuous, allows to understand the nature of the elementary excitations and to predict the
low-energy behavior of the materials (Goldstone modes, Mermin Wagner theorem, topological defects, ...). In some phases, at first glance,  the symmetry content may be hidden: as for example in Helium liquids.
The first obvious character is the absence of translation symmetry breaking and absence of a solid phase at zero temperature. It was very early understood (F.~London) that this absence of solidification
is due to the many-body quantum dynamics and the Helium phases have been named quantum liquids to be contrasted to the more ``classical liquids''. It was only decades after discovery of the  $^{4}${He}
superfluidity that the nature of the order parameter was unveiled. The understanding of the $^{3}${He} order parameter has also been heavily dependent on group symmetry considerations.

A parallel can be developed between this distinction of quantum liquids versus classical solids and that of spin liquids (SL's) versus N\'eel ordered phases. N\'eel ordered phases at least break translational symmetry of the
lattice and rotational symmetry of the spins. They can be described by a local order parameter and a Landau theory, whereas SL do not break any lattice symmetries nor spin rotation symmetry
and cannot be described by a \textit{local} order parameter. Similarly to $^{4}${He}, SL's can be characterized by an internal hidden, more or less complex order.

In this paper we  are mainly concerned with topological SL. These SL's are characterized at $T=0$ by exponentially decaying correlations for all  local observables (spins, dimers or spin nematic operators) and a spin gap to
bulk excitations.
They contrast to critical SL which have algebraic correlations and gapless excitations. It has been understood very early\cite{Read1989a,Kivelson1990} that the elementary excitations of these resonating valence bond (RVB) SL
carry a spin-$\frac{1}{2}$ contrarily to the spin-1 magnons of the N\'eel antiferromagnets. These emergent excitations are called spinons. A natural framework to describe the SL physics is the use of effective theories with the
fractional particles as elementary building blocks (parton construction). Going from the original spins to these fractionalized spinons implies the introduction of gauge fields in which the spinons are deconfined (SL) or glued (N\'eel order).
At first glance these approaches introduce via the gauge fields a considerable (infinite) amount of degrees of freedom.
In fact the number of possible distinct SL's is limited by the requirement that their physical observables do not break any lattice or spin symmetry and the enumeration of the different classes of distinct SL's can
be done through group theory analysis.

This was understood ten years ago by X.-G.~Wen who  developed a classification of symmetric SL using Projective Symmetry Group technique (PSG).\cite{Wen_PSG} The analysis of Wen for fermionic spinons on the square lattice was extended by
Wang and Vishwanath to bosonic spinons.\cite{PSG} In these  works, the definition of a SL is  limited to spin systems that do not break any symmetry, neither $SU(2) $ spin symmetry nor lattice symmetries nor time reversal symmetry.
These SL's have been dubbed by Wen \textit{symmetric} SL. This definition excludes \textit{chiral} SL which break time-reversal symmetry (and some minimal amount of lattice symmetry) but which do not break $SU(2)$ and do not have
long range order in spin-spin or dimer-dimer correlations.	

In the wake of Laughlin theory of FQHE, chiral SL's have been very popular at the end of the eighties (\textcite{Kalmeyer_Laughlin,Kalmeyer_89,Wen_Wilczek,Yang1993}), but, in the absence of indisputable candidates, this option has nearly disappeared from many discussions in the last decade.

Non-planar structures are quite  ubiquitous in classical frustrated magnetism,\cite{Regular_order}
and  are associated to scalar chirality: $\vec S_1\cdot(\vec S_2 \times \vec S_3)\ne 0$.
In some cases where the ground state is non-planar this   chirality can persists at finite temperature\cite{Momoi_classique,KagomeDomenge} although the magnetic order itself is absent for $T>0$ (Mermin-Wagner).
A similar phenomenon may take place in quantum systems at $T=0$.
There, the usual  scenario is that of a gradual reduction of the N\'eel order parameter when the strength of the quantum fluctuations is increased.
At some point the sublattice magnetization vanishes and  the $SU(2)$ symmetry is restored (leading to a SL).
Now, if the ordered magnetic structure is {\it chiral}, the time-reversal symmetry ${\cal T}$ may
still be  broken at the point where the magnetic order disappears, hence leading to a time-reversal symmetry breaking (TRSB) SL.\footnote{We will see in the following that in chiral phases, the fluxes can evolve continuously with the increase of quantum fluctuations, leading eventually to a non chiral SL phase. But this quantum phase transition has no reason to be concomitant with the opening of a (spinon) gap and the appearance of a SL,  as can be seen in Ref.~\onlinecite{cuboc1}.}
Some TRSB SL have indeed been recently proposed on the kagome lattice\cite{cuboc1,Faak2012} and  there are probably other examples.\cite{wen2010,chu2011}

The goal of this paper is to revisit the PSG analysis by relaxing the time-reversal symmetry constraint in order to include chiral SL's.
The framework used here is the Schwinger-boson mean-field theory (SBMFT).\footnote{SBMFT also coincides with large-$N$ limit of an $Sp(N)$ generalization of the $SU(2)$ model.\cite{ReadSachdev_SpN,Auerbach}}
But, as for the symmetric PSG, the symmetry considerations we use here should also be valid to classify SL in presence of moderate fluctuations beyond mean-field.

The paper is organized as follow.
Sections \ref{sec:recalls} and \ref{sec:SL} are reviews, to keep this article self-contained.
Section ~\ref{sec:recalls} is a description of SBMFT to fix the notations and precise the present understanding of this approach. Sec.~\ref{sec:SL} starts by recalling the gauge invariance of SBMFT and then describes how
the PSG is used to enforce the SL's symmetries on mean-field theories.

In Sec.~\ref{sec:LRO_CSL}, the concept of PSG is extended to include all chiral SL's.
In Sec.~\ref{sec:ansatze_tri} all the chiral and non chiral SL theories with explicit nearest neighbor gauge fields on the triangular lattice are derived. As an example of application  we propose a chiral SL as
the ground state of a ring-exchange model on the triangular lattice.
The physical meaning of the fluxes and their expressions in terms of spin operators is developed in Sec.~\ref{sec:fluxes}, as well as  the
question of topological loops on finite size samples.
 Sec.~\ref{sec:ccl} is the conclusion.
Appendices contain proofs of some statements in the main text, technical details and further applications to the square and kagome lattices.

\tableofcontents

\section{Schwinger boson mean-field theory (SBMFT)}
\label{sec:recalls}

We consider a spin Hamiltonian $\widehat H_0(\{\widehat{\mathbf S}_i\}_{i=1\dots N_s})$ on a periodic lattice with $N_s$ spins, each of length $S$. $\widehat H_0$ can contain Heisenberg interaction or more complicated terms
such as cyclic exchange, all invariant under global spin rotations ($SU(2)$ symmetry) and by time-reversal transformation $\cal T$
($\widehat H_0(\{\widehat{\mathbf S}_i\})=\widehat H_0(\{-\widehat{\mathbf S}_i\})$).
We insist on these symmetries since they are the basis of our construction.

Finding the ground state (GS) of a quantum spin problem is notoriously difficult problem and
the SBMFT provides an approximate way to treat the problem. This approach can be summarized by the following steps:
i) The spin operators (hence the Hamiltonian) are expressed using Schinwer bosons.
ii) A suitable rotationally-invariant mean-field decoupling leads to a quadratic Hamiltonian $H_{\rm MF}$.
iii)  $H_{\rm MF}$ is diagonalized using a Bogoliubov transformation and solved self-consistently.

\subsection{Bosonic operators and bond operators}

Let $m$ to be the number of sites per unit cell in the lattice, and $N_m$ number of unit-cells, so that $N_s=N_m m$ is the total number of sites.
We define the two bosonic operators $\widehat b_{i\sigma}^\dag$ that create a spin $\sigma=\pm1/2$ (or $\sigma=\uparrow$ or $\downarrow$) on site $i$.
The spin operators read:
\begin{subeqnarray}
\label{eq:def_S}
 \widehat S_i^z=\sum_\sigma\sigma\widehat b_{i\sigma}^\dag\widehat b_{i\sigma}, \\
 \widehat S_i^+=\widehat b_{i\uparrow}^\dag\widehat b_{i\downarrow}, \\
 \widehat S_i^-=\widehat b_{i\downarrow}^\dag\widehat b_{i\uparrow}.
\end{subeqnarray}
The Hamiltonian is thus a polynomial of bosonic operators with only even degree terms.
These relations imply that  the commutation relations
$[\widehat S^\alpha_i,\widehat S^\beta_i]=i\epsilon^{\alpha\beta\delta}\widehat S^\delta_i$
are verified. As for the total spin, it reads
$
 \vec{\widehat  S_i}^2=\frac{\widehat n_i}{2}\left(\frac{\widehat n_i}{2}+1\right)
$,
where $\widehat n_i=\widehat b^\dagger_{i\uparrow}\widehat b_{i\uparrow}
+\widehat b^\dagger_{i\downarrow}\widehat b_{i\downarrow}$ is the total number of bosons at site $i$.
To fix the ``length'' of the spins, the following constraint must therefore be imposed on physical states:
\begin{equation}
 \widehat n_i=\sum_\sigma\widehat b_{i\sigma}^\dag\widehat b_{i\sigma}=2S.
 \label{eq:constraint}
\end{equation}

In traditional MF theories, the MF parameter is the order parameter (as for example the magnetization $\langle \widehat{\mathbf S}_i\rangle$) and the MF Hamiltonian consequently breaks
the initial Hamiltonian symmetries, except in the high temperature phase where the MF parameter is zero.
Here, we would like to describe SL's that do not break any symmetry.
Thus we are going to express $\widehat H_0$ using quadratic bosonic operators, requiring their invariance by global spin rotations.

The expectation value  of these operators will then be used as mean-field parameters, insuring that the MF Hamiltonian  respects the  rotational invariance.
Only linear combinations of the two following operators and of their hermitian conjugates obey this property:
\begin{subeqnarray}
\widehat A_{ij}&=&\frac12(\widehat b_{i\uparrow}\widehat b_{j\downarrow}-\widehat b_{i\downarrow}\widehat b_{j\uparrow}),\\
\widehat B_{ij}&=&\frac12(\widehat b^\dag_{i\uparrow}\widehat b_{j\uparrow}+\widehat b^\dag_{i\downarrow}\widehat b_{j\downarrow}).
\end{subeqnarray}
$i$ and $j$ are lattice sites and these operators are thus bond operators.
They are linked by the relation
\begin{equation}
:\widehat B_{ij}^\dag \widehat B_{ij}:+\widehat A_{ij}^\dag \widehat A_{ij}=\frac14\widehat n_i(\widehat n_j -\delta_{ij})
\label{eq:relation_AB}
\end{equation}
where $:.:$ means normal ordering.

Any Hamiltonian invariant by global spin rotation can be expressed in terms of these operators only.
For example, an Heisenberg term $\widehat{\mathbf S}_i\cdot\widehat{\mathbf S}_j$ where $i\neq j$ can be decoupled as
\begin{subeqnarray}
\label{eq:SiSj}
\widehat{\mathbf S}_i\cdot\widehat{\mathbf S}_j
&=&:\widehat B_{ij}^\dag \widehat B_{ij}:-\widehat A_{ij}^\dag \widehat A_{ij},\slabel{eq:SiSj_a}
\slabel{eq:decoupling_AB}\\
&=&2:\widehat B_{ij}^\dag \widehat B_{ij}:-S^2,
\slabel{eq:decoupling_B}\\
&=&S^2-2\widehat A_{ij}^\dag \widehat A_{ij}.
\slabel{eq:decoupling_A}
\end{subeqnarray}
where the first line is true whatever the boson number, but the last two lines use Eq.~\ref{eq:relation_AB} and suppose that the constraint of Eq.~\ref{eq:constraint} is strictly respected.

To make clear the physical significance of these two bond operators in the case $S=\frac{1}{2}$, we write them
in terms of projection operators $\widehat P_s$ on the singlet state and $\widehat P_t$ on the triplet states:
\begin{subeqnarray}
\label{eq:projector}
 \widehat A_{ij}^\dag \widehat A_{ij}&=&\frac12 \widehat P_s\\
 :\widehat B_{ij}^\dag \widehat B_{ij}:&=&\frac14 (\widehat P_t-\widehat P_s).
\end{subeqnarray}
 We see in Eq.~\ref{eq:projector}, that $:\widehat B_{ij}^\dag \widehat B_{ij}:$ represents a ferromagnetic  contribution to Eq.~\ref{eq:decoupling_AB}, whereas $\widehat A_{ij}^\dag \widehat A_{ij}$ gives the singlet contribution.

\subsection{The mean-field approximation}

We now need two successive approximations to obtain a quadratic and solvable Hamiltonian.
We first relax the constraint on the boson number by imposing it only on {\it average}:
\begin{equation}
 \langle \widehat n_i\rangle =\kappa.
 \label{eq:constraint_mean}
\end{equation}
where $\kappa$ does not need to be a integer.
To implement this constraint, a Lagrange multiplier (or chemical potential) $\lambda_i$ is introduced at each site $i$ and the term $\sum_i \lambda_i(\kappa-\widehat n_i)$ is added to the Hamiltonian.
$\kappa$ can be continuously varied to interpolate between the classical limit ($\kappa=\infty$)  and the extreme quantum limit ($\kappa\to0$).

It should be reminded in general that fixing $\kappa=2S$ to study a spin-$S$ model is not necessarily the best choice as in the SBMFT
$\langle \widehat{\mathbf S}_{i}^2\rangle = \frac38 \kappa(\kappa+2)$.\cite{Auerbach}
An alternative choice could be to fix $\kappa$ in such a way that the spin fluctuations and not the spin length have the correct value.\footnote{The
SBMFT is not meant to provide an accurate quantitative agreement with $SU(2)$ models, nor to replace controlled numerics.
SBMFT should instead be viewed as a tool to identify possible phases in competition, their instabilities, to provide indicative phase diagrams, and to identify the important degrees of freedom that
need to be incorporated in a description going beyond mean field.}

In a second step, bond operators fluctuations are neglected and a MF Hamiltonian $\widehat H_{\textrm {MF}}$ that is linear in bond operators is obtained.
For instance:
\begin{equation}
 :\widehat B_{ij}^\dag \widehat B_{ij}:\,\simeq \langle \widehat B_{ij}^\dag\rangle \widehat B_{ij}+ \widehat B_{ij}^\dag\langle \widehat B_{ij}\rangle -|\langle \widehat B_{ij}\rangle|^2.
\end{equation}
We replace $\langle \widehat B_{ij}\rangle$ and $\langle \widehat A_{ij}\rangle$ by complex bond parameters $\mathcal A_{ij}$ and $\mathcal B_{ij}$.
This MF approximation can be seen as the first term of a large $N$ expansion of a $Sp(N)$ theory.\cite{ReadSachdev_SpN}
The steps are explained in details in Ref.~\onlinecite{Auerbach} in the very similar case of an $SU(N)$ theory.
This zero'th order $1/N$ expansion can be pursued to the first order.\cite{Trumper_fluct}
The MF Hamiltonian is now a quadratic bosonic operator.
It can be written in terms of a $2N_s\times 2N_s$ complex matrix $M$ and of a real number $\epsilon_0$ depending on the $\mathcal A_{ij}$ and $\mathcal B_{ij}$ and on the Lagrange multipliers $\lambda_i$:
\begin{equation}
\label{eq:HMF}
 \widehat H_{\textrm {MF}}=\phi^\dag M \phi+\epsilon_0.
\end{equation}
where $\phi^\dag=(\widehat b_{1\uparrow}^\dag,\widehat b^\dag_{2\uparrow},\dots,\widehat b^\dag_{N_s\uparrow},\widehat b_{1\downarrow},\dots,\widehat b_{N_s\downarrow})$.\footnote{The most general quadratic
Hamiltonian would require the use of a larger vector
$(\widehat b_{1\uparrow}^\dag,\dots,\widehat b^\dag_{N_s\uparrow},\widehat b_{1\downarrow}^\dag,\dots,\widehat b^\dag_{N_s\downarrow},\widehat b_{1\uparrow},\dots,\widehat b_{N_s\downarrow})$
but the rotational invariance implies conservation of $\hat S^z_{\rm tot}=\sum_i \widehat b^\dagger_{i\uparrow}\widehat b_{i\uparrow}
+\widehat b^\dagger_{i\downarrow}\widehat b_{i\downarrow}$ and the allowed quadratic terms are therefore limited to
$\widehat b_{i\uparrow}^\dag\widehat b_{j\uparrow}$, $\widehat b_{i\downarrow}\widehat b_{j\downarrow}^\dag$, $\widehat b_{i\uparrow}^\dag\widehat b_{j\downarrow}^\dag$ and $\widehat b_{i\downarrow}\widehat b_{j\uparrow}$.}
The expression for $M$ and  $\epsilon_0$ depend on $\widehat H_0$ and on the chosen decoupling (for example using Eq.~\ref{eq:decoupling_AB}, \ref{eq:decoupling_B} or \ref{eq:decoupling_A}).

The set of mean-field parameters $\{\mathcal A_{ij},\mathcal B_{ij}\}$ appearing in $H_\textrm {MF}$ is called an \textbf{Ansatz}.
Up to an equivalence relation that will be described in the next section, an Ansatz defines a specific phase (ground state and excitations).
Depending on the value of $\kappa$, this state can either have N\'eel long range order,
or the bosons are gapped (several types of SL are then possible).

In the following we will explain and exploit the relation which exists between regular classical magnetic orders\cite{Regular_order} and SL's.

To enforce self-consistency, the following conditions should be obeyed:
\begin{equation}
 \mathcal A_{ij}=\langle \widehat A_{ij}\rangle \,\textrm{ and }\, \mathcal B_{ij}=\langle \widehat B_{ij}\rangle,
\end{equation}
which are equivalent to
\begin{equation}
\label{eq:self_cons2}
 \frac{\partial F_{MF}}{\partial \mathcal A_{ij}}=0 \,\textrm{ and }\, \frac{\partial F_{MF}}{\partial \mathcal B_{ij}}=0,
\end{equation}
where $F_{MF}$ is the MF free energy, together with the constraint
\begin{equation}
\label{eq:constraint3}
 \langle \widehat n_{i}\rangle=\kappa\,\Leftrightarrow\,\frac{\partial F_{MF}}{\partial \mathcal \lambda_{i}}=0.
\end{equation}
The next step is to calculate the mean values of the operators $\widehat A_{ij}$ and $\widehat B_{ij}$ either in the GS of $\widehat H_{\textrm {MF}}$ if the temperature is zero,
or in the equilibrium state for non-zero temperatures.
In both cases one needs to use a Bogoliubov transformation to diagonalize $H_{\rm MF}$.
As this transformation is often explained in the simple case of $2\times2$ matrices (or for particular sparse matrices),
we explain the algorithm in a completely general case in App.~\ref{app:bogoliubov}.

\subsection{Choice of bond fields: \texorpdfstring{$\widehat A_{ij}$}{} and \texorpdfstring{$\widehat B_{ij}$}{} or \texorpdfstring{$\widehat A_{ij}$}{} or \texorpdfstring{$\widehat B_{ij}$}{} only.}

As in the example of Eq.~\ref{eq:SiSj}, the relation~\ref{eq:relation_AB} can be used to
eliminate $\mathcal A_{ij}$ or $\mathcal B_{ij}$ from $\widehat H_{\textrm {MF}}$.
If we choose to keep only the $\mathcal B_{ij}$ parameters, $M$ is block diagonal with two blocks of size $N_s$ and the vacuum of bosons is a GS.
To obey the constraint of Eq.~\ref{eq:constraint}, we have to adjust the boson densities by filling some zero-energy mode(s), therefore breaking the $SO(3)$ symmetry.
The GS is thus completely classical.
On the contrary, we can keep the $\mathcal A_{ij}$ only, but then the singlet weight is overestimated, which can be introduce some bias on frustrated lattices where short-distance correlations are not collinear.
Keeping $\widehat A_{ij}$ only is a widespread practice in the litterature, but Trumper {\it et al.}\cite{Trumper_AB_SBMFT} have explicitly shown that the bandwidth of the spectrum of excitations of the Heisenberg model on the triangular lattice
is twice too large when using $\mathcal A_{ij}$ fields only. On the other hand, the use of both $\mathcal A_{ij}$ and $\mathcal B_{ij}$ restores the correct bandwidth and a improves quantitatively the excitation spectrum.
Note that even on the square lattice the simultaneous use of both bond operators improves the ground state energy.\cite{CecattoSquare}

From a different point of view Flint and Coleman\cite{Symplectic_SBMFT} advise the use of both fields in order to have a large-$N$ limit where spin generators are {\it odd} under the time-reversal symmetry, as it is the case for $SU(2)$.

\section{The search of SL}
\label{sec:SL}

Even when considering an Hamiltonian with nearest neighbor interactions only, the dimension of the MF parameters manifold is exponentially large.\footnote{
With a coordination number $z$ and two complex parameters per bond, it is naively $\mathbb C^{zN_s}$.
In fact, the modulus of self-consistent $\widehat A_{ij}$ and $\widehat B_{ij}$ cannot exceed a
bound, which depends on $\kappa$, see the details in App.~\ref{app:borne}. As for the gauge invariance, it allows to fix some parameters to be real, see below.}
Moreover the Lagrange multipliers $\lambda_i$ make the search of the stationary points of the MF free energy difficult (constrained optimization) as for each considered Ansatz, all $\lambda_i$
must be adjusted to calculate the MF free energy.
In Ref.~\onlinecite{SBMFT_alllinks} this optimization  was carried out (without any simplifying/symmetry assumption) on square and triangular lattices with up to 36 sites. In almost all cases the MF ground-state turned
out to be highly symmetric, as expected, but  excited mean-field solutions are however highly inhomogeneous (and often not understood yet).
The problem can be considerably simplified if we restrict our search to states respecting some (or all) the symmetries of $\widehat H_0$.
Such symmetries are divided into global spin rotations, lattice symmetries and time reversal symmetry.
We have assumed from the beginning that $\widehat H_0$ is invariant by  global spin rotations and chosen the MF approximation in such a way that it remains true for $\widehat H_{\textrm {MF}}$,
but the choice of a specific  Ansatz may  or may not break other discrete symmetries.
The fore-coming section explains how to find all Ans\"atze such as the physical quantities are invariant by all the lattice symmetries $\mathcal X$, either strictly (for symmetric SL's) or only up to a time-reversal transformation (Chiral SL's).

We will now define some groups specific to an Ansatz: the invariance gauge group  in Sec.~\ref{sec:IGG}  and the projective symmetry group in Sec.~\ref{sec:PSG}.
Then, in Sec.~\ref{sec:APSG}, we define the  algebraic projective symmetry group, which is associated to a lattice symmetry group and not specific to a particular  Ansatz on this lattice.

\subsection{Gauge invariance, fluxes and invariance gauge goup (IGG)}
\label{sec:IGG}

Let $\mathcal G \backsimeq U(1)^{N_s}$ be the set of gauge transformations.
A  gauge transformation is characterized by an angle $\theta(i) \in[0,2\pi[$ at each site and the operator $\widehat G$ which implements the associated gauge transformation
\begin{equation}
\label{eq:gauge_transf}
 \widehat b_{j\sigma}\to \widehat b_{j\sigma}\,e^{i\theta(j)} = \widehat G^\dagger  \widehat b_{j\sigma}\widehat G
\end{equation}
is given by
\begin{equation}
 \widehat G=\exp\left(i\sum_j \widehat b_{j\sigma}^\dagger \widehat b_{j\sigma} \theta(j)\right).
\end{equation}
A wave function $|\phi\rangle$ respects a symmetry $\widehat F$ if all the physical observables measured in the state $\widehat F|\phi\rangle$ are identical to those measured in $|\phi\rangle$.
It does not mean that $|\phi\rangle=\widehat F|\phi\rangle$, but that the two wave functions are equal up to a gauge transformation: $\exists\, \widehat G\in\mathcal G, |\phi\rangle=\widehat G\widehat F|\phi\rangle$.

The action of $\widehat G$ on the Ansatz is:
\begin{equation}
\left\{
\begin{array}{l}
 \mathcal A_{jk}\to \mathcal A_{jk}\,e^{i(\theta(j)+\theta(k))},\\
 \mathcal B_{jk}\to \mathcal B_{jk}\,e^{i(-\theta(j)+\theta(k))},
\end{array}\right.
\end{equation}
such as $\widehat H_{\textrm {MF}}$ remains unaffected by $\widehat G$.
We note that $\langle \widehat A_{jk}\rangle$ and $\langle \widehat B_{jk}\rangle$ are gauge dependent: they are not physical quantities as they do not preserve the on-site boson number.
As any such quantity, their mean values calculated using $\widehat H_0$ is zero when the average is taken on all gauge choices.
Using $\widehat H_{\textrm {MF}}$, it can be non-zero as the gauge symmetry is explicitly broken by the choice of the Ansatz.

We have seen that changing the gauge modifies the Ansatz but not the physical quantities.
Conversely, if two MF Hamiltonians give rise to the same physical quantities, then their Ans\"atze are linked by a gauge transformation.
In fact two types of physical quantities are directly related to the Ansatz: the MF parameter moduli (related to the scalar product of two spins), and the fluxes. The fluxes are defined as the arguments of Wilson
loop operators such as $\langle \widehat B_{ij}\widehat B_{jk}\widehat B_{ki}\rangle$ or of $\langle \widehat A^\dag_{ij}\widehat A_{jk}\widehat A^\dag_{kl}\widehat A_{li}\rangle$. By construction these quantities are
gauge invariant and define the Ansatz up to gauge transformations.
The physical meaning of fluxes will be addressed in Sec.~\ref{sec:fluxes}.

The gauge transformations that do not modify a specific Ansatz form a subgroup of $\mathcal G$ called the invariance gauge group (IGG).
It always contains the minimal group $\mathbb Z_2$ formed by the identity and by the transformation Eq.~\ref{eq:gauge_transf} with $\theta(i)=\pi$ for all lattice sites $i$.
In the particular cases where we can divide the lattice in two sublattices such as $\mathcal A_{ij}=0$ whenever $i$ and $j$ are in the same sublattice (bipartite problem), the IGG is enlarged to $U(1)$.
The later situation corresponds, for instance, to an Ansatz on a square lattice with only first-neighbor $\mathcal A_{ij}$.
The transformations of the IGG are then given by $\theta(i)=\theta$ on a one sublattice and $\theta(i)=-\theta$ on the other, with arbitrary  $\theta\in[0,2\pi[$.

\subsection{The projective symmetry group (PSG)}
\label{sec:PSG}

Let $\mathcal X$ be the group of the lattice symmetries of the Hamiltonian $\widehat H_0$ (translations, rotations, reflections\dots).
From now on, for the sake of simplicity, we discard the hat on the gauge and symmetry operators.
The effect of an element $X$ of $\mathcal X$ on the bosonic operators is
\begin{equation}
\label{eq:latt_sym}
 X:\widehat b_{j\sigma}\to \widehat b_{X(j)\sigma}.
\end{equation}
The effect of $X$ on the Ansatz is:
\begin{equation}
\left\{
\begin{array}{l}
 \mathcal A_{jk}\to \mathcal A_{X(j)X(k)},\\
 \mathcal B_{jk}\to \mathcal B_{X(j)X(k)}.
\end{array}\right.
\end{equation}
We know that a gauge transformation does not change any physical quantities.
What about the lattice symmetries ?
We know from Sec.~\ref{sec:IGG} that if the Ans\"atze before and after the action of $X$ have the same physical quantities, they are linked by a gauge transformation: it thus exists at least one gauge transformation
$G_X$ such as $G_XX$ leaves the Ansatz unchanged.
\textit{The set of such transformations of $\mathcal G\times \mathcal X$ is called the projective symmetry group (PSG) of this Ansatz.}
Note that this group only depends on the Ansatz and on $\mathcal X$, but not on the details of the Hamiltonian.
Thus, an Ansatz is said to respect a lattice symmetry $X$ if it exists a transformation $G_X \in \mathcal G$ such that the Ansatz is invariant by $G_XX$.

The IGG of an Ansatz is the PSG subgroup formed by the set of gauge transformations $G_I$ associated to the identity transformation $I$ of $\mathcal X$.
For each lattice symmetry $X \in \mathcal X$ respected by the Ansatz, the set of gauge transformations $G_X$ such as $G_XX$ is in the PSG is isomorph to the IGG: for any $G_I$ in the IGG, $(G_IG_X)X$ is in the PSG.
Thus, the condition for an Ansatz to respect all the lattice symmetries is that its PSG is isomorph to IGG$\times\mathcal X$.

\subsection{The algebraic projective symmetry groups}
\label{sec:APSG}

An Ansatz is characterised (partially) by its IGG and its PSG. In turn, we know from these groups which lattice symmetries it preserves.
Reversely we now want to impose lattice symmetries and find all Ans\"atze that preserves them.
To reach this goal, we proceed in two steps.
The first one is to find the set of the so-called algebraic PSG's. \cite{Wen_PSG,PSG}
They are subgroups of $\mathcal G\times \mathcal X$ verifying algebraic conditions necessarily obeyed by a PSG.
Contrary to the  PSG of an Ansatz, the algebraic PSG's exist independently of any Ansatz and only depend on the lattice symmetry group $\mathcal X$ and on the choice of an  IGG (chosen as the more general).
An algebraic PSG does not depend on the details of the lattice such as the positions of the sites.
However, depending on these details, an algebraic PSG may  have zero, one, or many compatible Ans\"atze.
The second step consists, for a given lattice,  in finding all the Ans\"atze  compatible with a given algebraic PSG.

Let us detail the algebraic conditions verified by the algebraic PSG's.
The group $\mathcal X$ is characterized by its generators $x_1$\dots$x_p$.
A generator $x_a$ has an order $n_a\in \mathbb N^*$ such as $x_a^{n_a}$ is the identity
(if no such integer exists, we set $n_a=\infty$).
For any transformation $X\in\mathcal X$, there exists a unique ordered product $X=x_1^{k_1}\dots x_p^{k_p}$ with $0\leq k_a<n_a$ if $n_a$ is finite, $k_a\in\mathbb Z$ if not.
The rules used to transform an unordered product into an ordered one are the algebraic relations of the group.
Each of these rules implies a constraint on the $G_{x_a}$ (chosen as one of the gauge transformation associated to $x_a$).
Basically, it states that if a lattice symmetry $X$ can be written in several ways using the generators, the gauge transformation $G_X$ is independent of the writing (up to an IGG transformation).
The subgroups of $\mathcal G\times \mathcal X$ respecting all these constraints are the algebraic PSG's.

To illustrate the idea, let us consider a basic example where $\mathcal X$ is generated by two translations $x_1$ and $x_2$.
Both transformations have an infinite order $n_1=n_2=\infty$.
We have $X\in \mathcal X$ written as product of generators $X=x_1^{m_1}x_2^{m_2}x_1^{m_3}x_2^{m_4}\dots$ and we would like o write it as $X=x_1^{p_1}x_2^{p_2}$.
The  need algebraic relation is simply the commutation between the two translations : $x_1x_2=x_2x_1$.
We then have $p_1=m_1+m_3+\dots$ and $p_2=m_2+m_4+\dots$.
We will now see that this implies a constraint on $G_{x_1}$ and $G_{x_2}$.
Suppose that we have an Ansatz unchanged by $G_{x_1}x_1$ and $G_{x_2}x_2$.
Then the inverses $x_1^{-1}G_{x_1}^{-1}$ or $x_2^{-1}G_{x_2}^{-1}$ too are in the PSG.
So, the product $G_{x_1}x_1G_{x_2}x_2x_1^{-1}G_{x_1}^{-1}x_2^{-1}G_{x_2}^{-1} \in$ PSG.
This product has been chosen to make the algebraic relation $x_1x_2=x_2x_1$ ($\Leftrightarrow x_1x_2x_1^{-1}x_2^{-1}=I$) appear after the following manipulations:
\begin{eqnarray*}
& G_{x_1}x_1G_{x_2}x_2x_1^{-1}G_{x_1}^{-1}x_2^{-1}G_{x_2}^{-1} \in \rm{PSG}\\
\Leftrightarrow&G_{x_1}(x_1G_{x_2}x_1^{-1})x_1x_2x_1^{-1}x_2^{-1}(x_2G_{x_1}^{-1}x_2^{-1})G_{x_2}^{-1} \in \rm{PSG}\\
\Leftrightarrow&G_{x_1}(x_1G_{x_2}x_1^{-1})(x_2G_{x_1}^{-1}x_2^{-1})G_{x_2}^{-1} \in \rm{PSG}.
\end{eqnarray*}
The expressions in parenthesis in the last line are pure gauge transformations and the full resulting expression is a product of gauge transformations.
Thus, we can more precisely write:
\begin{equation}
 G_{x_1}(x_1G_{x_2}x_1^{-1})(x_2G_{x_1}^{-1}x_2^{-1})G_{x_2}^{-1} \in \rm{IGG}.
\end{equation}
If the IGG is $\mathbb Z_2$, this constraint can be written in term of the phases $\theta_X(i)$ of the gauge transformation $G_X$ as:
\begin{equation}
 \theta_{x_1}(i)+\theta_{x_2}(x_1^{-1}i)-\theta_{x_1}(x_2^{-1}i)-\theta_{x_2}(i)=p\pi,
\end{equation}
with $p=0$ or $1$.
This constraint coming from the commutation relation between $x_1$ and $x_2$ must be obeyed by all algebraic PSG's.

It is useless to list all algebraic PSG's for the simple reason that some of them are equivalent and give Ans\"atze with the same physical observables.
Two (algebraic or not) PSG's are equivalent if they are related by a gauge transformation $G$: for any gauge transformation $G_X$ associated to the lattice symmetry $X$ in the first PSG, $GG_XG^{-1}$ belongs to the set of gauge transformations associated to $X$ in the second PSG.
We are only interested in equivalence classes of PSG's.

Taking algebraic PSG's in different classes does not imply that they have no common Ans\"atze: a trivial example is the Ansatz with only zero parameters, belonging to any algebraic PSG's.
But each class includes Ans\"atze that are in no other class and have specific physical properties.

Once all the algebraic PSG's classes are determined, it remains to find the possible compatible Ans\"atze for one representant of each class.
As an example of compatibility condition, let's take the case where $X$ belongs to the considered algebraic PSG ({\it i.e.} $G_{X}=I$).
Then an Ansatz can be compatible with this algebraic PSG only if, for any couple of sites $(i,j)$, $\mathcal A_{ij}=\mathcal A_{X(i)X(j)}$.
If such compatible Ans\"atze exist, they respect the lattice symmetries by construction (in the sense that their physical quantities do so).
We now want to impose the time reversal symmetry: among the compatible Ans\"atze, we only keep those that are equivalent to a real Ansatz up to a gauge transformation.
We call them \textit{strictly symmetric} Ans\"atze (\textit{weakly symmetric} ones are defined in the next section).

To completely define an Ansatz, it is sufficient to give the algebraic PSG and the values of the MF parameters on non symmetry-equivalent bonds.
For example, on a square (or triangular or kagome) lattice with all usual symmetries (see Fig.~\ref{fig:sym_latt}) and only first neighbor interactions, the $\mathcal A_{ij}$ and $\mathcal B_{ij}$ of one bond are enough.

\section{From chiral long range orders to chiral SL's}
\label{sec:LRO_CSL}

We will now show  that the zoo of N\'eel LRO obtained from the strictly symmetric Ans\"atze misses
the chiral states which are exact ground states of a large number of frustrated classical models. This will lead us in a straightforward manner to the construction of chiral algebraic PSG's in which time reversal and some lattice symmetries can be broken (Sec.~\ref{sec:CSL}). This generalised framework will then be illustrated on the triangular lattice in Sec.~\ref{sec:ansatze_tri} and on the square and kagome lattice in App.~\ref{app:EAPSG}.

\subsection{\texorpdfstring{$SU(2)$}{} symmetry breaking of symmetric Ans\"azte}
\label{sec:classical_order}

To simplify, we suppose that all lattice sites are equivalent by symmetry and only consider Ans\"atze such as the $\lambda_i$ are all equal to a single $\lambda$.
Even if an Ansatz is strictly symmetric, it does not always represent a SL phase.
As is well known in SBMFT, a Bose condensation of zero energy spinons can occur and leads to N\'eel order.
We will discuss how the Ans\"azte symmetry constraints the magnetic order obtained after condensation, and establish a relation with the \textit{regular states} introduced in Ref.~\onlinecite{Regular_order}.

The Bogoliubov bosons creation operators are linear combinations of the $\widehat b_{i\sigma}$ and $\widehat b_{i\sigma}^\dag$, such as their vacuum $|\tilde 0\rangle$ is a GS of
$\widehat H_{\rm{MF}}$ (see App.~\ref{app:bogoliubov}).
If the GS is unique, it must respect all the Hamiltonian symmetries and consequently, cannot break the global spin rotation invariance.
But when $\kappa$ increases (we continuously adapt the Ansatz to $\kappa$ so that the self-consistency conditions remains verified and the PSG remains the same), some eigen energie(s) decrease(s) to zero.
The GS is then no more unique as the zero mode(s) can be more or less populated and the phases of each zero mode are free.
It is then possible to develop a long range spin order.

This phenomena occurs when no $\lambda$ verifies condition \ref{eq:constraint3}.
If $\lambda$ increases the mean number of boson per site increases up to a maximal number $\kappa_{\rm{max}}$.
At this point, some eigen energies become zero.
Increasing $\lambda$ further is not possible  as the Bogoliubov transformation becomes unrealizable (the $M$ matrix of Eq.~\ref{eq:HMF} has non-positive eigenvalues).
To reach the required number of boson per site, we have to fill the zero energy modes ${\tilde b}^\dag_1$, ${\tilde b}^\dag_2$, \dots using
coherent states $e^{\alpha_1{\tilde b}^\dag_1+\alpha_2{\tilde b}^\dag_2+\dots}|\tilde 0\rangle$ for example.
In the thermodynamical limit the fraction of missing bosons is macroscopic and a Bose condensation occurs in each of the soft modes.
The choice of the weight $\alpha_i$ of these modes fixes the direction of the on-site magnetization.
Detailed examples of magnetization calculations in a condensate are given by Sachdev.\cite{Sachdev}

In the classical limit ($\kappa\to\infty$), all bosons are in the condensate and contribute to the on-site magnetization $\mathbf m_i$.
The  modulus $|\mathbf m_i|$  should be equal to $\kappa/2$ to satisfy Eq.~\ref{eq:HMF}.
The $\widehat b_{i\sigma}$ operators acquire a non-zero expectation value $\langle \widehat b_{i\sigma}\rangle$ and are (up to a gauge transformation) linked to $\mathbf m_i$ by :
\begin{equation}
\left(
\begin{array}{c}
\langle \widehat b_{i\uparrow}\rangle\\
\langle \widehat b_{i\downarrow}\rangle
\end{array}\right)=
 \left(
\begin{array}{c}
\sqrt{|\mathbf m_i|+ m_i^z}\\
\sqrt{|\mathbf m_i|- m_i^z}e^{i{\rm Arg}(m_i^x+i m_i^y)}
\end{array}\right),
\label{eq:ab_class}
\end{equation}
where Arg is the argument of the complex number and $m_i^{x,y,z}$ are the magnetization components.
These values are constrained by the Ansatz through:
\begin{subeqnarray}
\label{eq:AB_class}
\mathcal{A}_{ij}
&=&\frac{1}{2}(\langle \widehat b_{i\uparrow}\rangle\langle \widehat b_{j\downarrow}\rangle-\langle \widehat b_{j\uparrow}\rangle\langle \widehat b_{i\downarrow}\rangle),
\\
\mathcal{B}_{ij}
&=&\frac{1}{2}(\langle \widehat b^\dag_{i\uparrow}\rangle\langle \widehat b_{j\uparrow}\rangle+\langle \widehat b^\dag_{i\downarrow}\rangle\langle \widehat b_{j\downarrow}\rangle).
\end{subeqnarray}
The supplementary constraint reads:
\begin{equation}
\label{eq:constraint_class}
 |\mathbf m_i|\sim \kappa/2
\end{equation}
This extra constraint can make the classical limit problem unsolvable: no  classical magnetization pattern is then compatible with the Ansatz.
An example of such a situation was studied by \textcite{PSG} (see App.~\ref{app:strange}).

We can take the problem of the classical limit from the other side.
We begin from a classical state, from which we calculate $\langle\widehat b_{i\sigma}\rangle$ and the Ansatz (using Eq.~\ref{eq:ab_class} and \ref{eq:AB_class}).
What are the conditions on the classical state for the associated Ansatz to be strictly symmetric ?
As we look for an Ansatz respecting all lattice symmetries, the rotationally invariant quantities (as the spin-spin correlations) must be invariant by all lattice symmetries, what severely limits the classical magnetization pattern.
Such a state is called a $SO(3)$-regular state.
Mathematically,  a state is said to be $SO(3)$-regular if for any lattice symmetry $X$ there is a global spin rotation $S_X\in SO(3)$ such as the state is invariant by $S_XX$.
Moreover, the time reversal symmetry ({\it i.e.} the Ansatz can be chosen to be real) imposes the co-planarity of the spins.\footnote{If we choose the $xz$ plane, we directly obtain a real Ansatz from Eqs.~\ref{eq:ab_class},\ref{eq:AB_class}.}
The set of coplanar $SO(3)$-regular states can be sent on the set of condensed states of strictly symmetric Ans\"atze.
In the same way, we define the $O(3)$-regular states by including global spin flips $\mathbf S_i\to-\mathbf S_i$ in the group of spin transformations.
These $O(3)$-regular states are listed in Ref.~\onlinecite{Regular_order} for several two-dimensional  lattices.
The $O(3)$-regular states are divided in coplanar $SO(3)$-regular states and in chiral states.
In a chiral state, the global inversion $\mathbf S_i\to-\mathbf S_i$ cannot be ``undone'' by a global spin rotation. Equivalently, there exist three sites
$i$, $j$, $k$ such as the scalar chirality $\mathbf S_i\cdot (\mathbf S_j\land \mathbf S_k)$ is non zero: the spins are not coplanar.
Then a strictly symmetric Ansatz, upon condensation, can only give coplanar $SO(3)$-regular states in the classical limit, therefore missing all chiral $O(3)$-regular states.

\begin{figure}
\begin{center}
 \includegraphics[height=.14\textwidth]{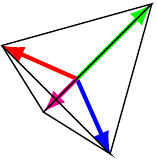}\quad
 \includegraphics[height=.14\textwidth]{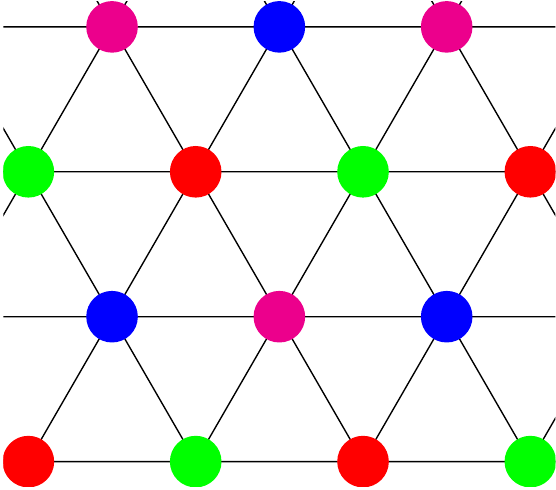}
 \caption{(Color online) Tetrahedral order on the triangular lattice}
 \label{fig:order_tri_tetra}
\end{center}
\end{figure}

This limitation can seem unimportant as most of the usual long-range ordered spin models have  planar GS's.
But some new counter examples have recently been discovered.
The first example is the cyclic exchange model on the triangular lattice\cite{Momoi_classique} with a four sublattice tetrahedral chiral GS (see Fig.~\ref{fig:order_tri_tetra}).
More recently,  two twelve sublattice chiral GS's, with the spins oriented towards the corners of a cuboctahedron, were discovered on the kagome lattice with first and second neighbor exchanges\cite{KagomeDomenge,Janson_2008} (studied in App.~\ref{app:kagome_ansatz}).
A systematic study of the classical GS's of simple models on different lattices has indeed revealed that the GS's are chiral for large ranges of interaction values.\cite{Regular_order}

The theory of symmetric PSG is unable to encompass such chiral states.
In the following subsection, we will build TRSB SL Ans\"atze which include, upon condensation, all classical regular chiral states.
This method was already applied to the kagome lattice with up to third neighbor interactions, leading to the surprising result of a chiral state even in the purely first neighbor model. \cite{cuboc1}
If this state is physically relevant or not is still an open question, but independently, it shows that the 	
omission of chiral Ans\"atze has prevented the discovery of more competitive MF solutions.

\subsection{The chiral algebraic PSG's: how to include weakly symmetric states}
\label{sec:CSL}

The time-reversal transformation $ \cal T$ acts on an Ansatz by complex conjugation of the MF parameters.\cite{Wen_PSG}
If an Ansatz respects this symmetry, it is sent to itself by $\cal T$ (up to a gauge transformation).
So, in an appropriate gauge, all parameters can then be chosen real.
In most previous SBMFT studies, the hypothesis of time reversal invariance of the GS was implicit, as only \textit{real} Ans\"atze were considered.
In contrast to $SU(2)$ global spin symmetry that can easily be broken through the Bose condensation process,
no transition is known to produce a chiral ordered state out of a $ \cal T$-symmetric Ansatz.
Indeed, chiral Ansätze have loops with complex-valued fluxes  which  evolve continuously with $\kappa$.
We do not expect any singular behavior of these (local) fluxes when crossing the condensation point, so the generic situation
is that a chiral LRO phase will give rise to  a TRSB SL\cite{cuboc1,Faak2012} when decreasing $\kappa$.
It is of course possible that the lowest-energy Ansatz changes with $\kappa$ but such a first-order transition has not reason to coincide with the onset of magnetic LRO.

To obtain all chiral SL's we have to explicitly break time-reversal symmetry at the MF level, in the Ansatz.
For $SO(3)$ classical regular states, a lattice transformation from $\mathcal X$ is compensated by a global spin rotation (that leaves the Ansatz unchanged).
For $O(3)$ classical regular states, a lattice transformation $X\in\mathcal X$ is compensated by a global spin rotation possibly followed by an inversion $\mathbf S_i\to-\mathbf S_i$.
This defines a parity $\epsilon_X$ to be  $+1$ if no spin inversion is needed, and $-1$ otherwise.
In a chiral SL, the parity will be deduced from the effect of $X$ on the fluxes: $\epsilon_X=1$ if they are unchanged, $-1$ if they are reversed.
With this distinction in mind we will call \textit{weakly symmetric} Ans\"atze (WS)
the Ans\"atze respecting the lattice symmetries up to $ \cal T$, whereas the the Ans\"atze respecting strictly all lattice symmetries and $\cal T$ have already been called
\textit{strictly symmetric} (SS) Ans\"atze (all lattice symmetries are even).

The distinction between even and odd lattice symmetries (as defined by $\epsilon_X$) is the basis of the construction of all WS Ans\"atze via the chiral algebraic PSG's.
Let us consider $\mathcal X_e$ the subgroup of transformations of $\mathcal X$ that can only be even.
Mathematically, $\mathcal X_e$ is the subgroup of $\mathcal X$ which elements are sent to the identity by {\it all} morphisms from $\mathcal X$ to $\mathbb Z_2$.
$\mathcal X_e$ contains at least all the squares of the elements of $\mathcal X$ as $\epsilon_{X^2}=\epsilon_X^2=1$.
But, depending on the algebraic relations of $\mathcal X$, it may contain more transformations as we show in the triangular case in Sec.~\ref{sec:APSG_tri}.
Once $\mathcal X_e$ is known, we define the {\it chiral} algebraic PSG's of $\mathcal X$ as the algebraic PSG's of $\mathcal X_e$.
The method described previously to find all algebraic PSG's applies the same way.
We define $\mathcal X_o$ as the set  of transformations which may be odd ($\mathcal X-\mathcal X_e$). It contains transformations of undetermined parities.

To filter the weakly symmetric Ans\"atze from those compatible with the chiral algebraic PSG's, we have to take care of the transformations of $\mathcal X_o$.
This gives two types of extra constraints.
First, same type ($\mathcal A$ or $\mathcal B$) MF parameters on bonds linked by such transformation must have the same modulus.
The second constraint concerns their phases, through the fluxes.
The phases are gauge dependent, but the fluxes are gauge independent.
Fluxes are sent to their opposite by $\cal T$ and as well as by the odd transformations of $\mathcal X$. Then are unchanged by even transformations.
To find all WS Ans\"atze we then have to determine a maximal set of independent elementary fluxes and distinguish all possible cases of parities for the transformations of $\mathcal X_o$ ($\epsilon_X=\pm1$).

We can now apply these theoretical considerations to find all WS Ans\"atze on some usual lattices as the triangular, honeycomb, kagome and square lattice.
The calculations are detailed for the triangular lattice in the following subsections and some results for the kagome and square lattice are given in App.~\ref{app:EAPSG}.

\subsection{Chiral algebraic PSG's of lattices with a triangular Bravais lattice}
\label{sec:APSG_tri}

\begin{figure}
\begin{center}
 \includegraphics[height=.19\textwidth]{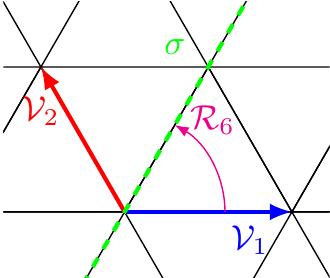}\quad
 \includegraphics[height=.19\textwidth]{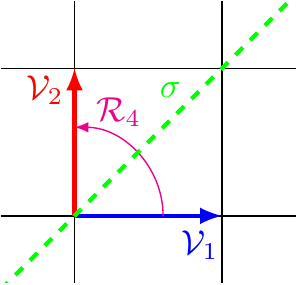}\\
 \caption{(Color online) Generators of the lattice symmetries $\mathcal X$ on the triangular and square lattices.
$\mathcal V_i$ is a translation, $\sigma$ is a reflection and $\mathcal R_i$ is a rotation of order $i$. }
 \label{fig:sym_latt}
\end{center}
\end{figure}

The first step is to find all chiral algebraic PSG's.
As already mentioned, they only depend on the symmetries of $\mathcal X_e$ and on the IGG.
We choose the most general case of IGG$\sim\mathbb Z_2$ and suppose that $\widehat H_0$ respects all the lattice symmetries with the generators described in Fig.~\ref{fig:sym_latt}.
These symmetries are those of a triangular lattice, but the actual (spin) lattice of $\widehat H_0$ can be any lattice with a triangular Bravais lattice such as a honeycomb, a kagome or more complex lattices.
The coordinates $(x,y)$ of a point are given in the basis of the translation vectors $\mathcal V_1$, $\mathcal V_2$ and the effect of the generators on the coordinates are
\begin{subeqnarray}
 \mathcal V_1:(x,y)&\to&(x+1,y),\\
 \mathcal V_2:(x,y)&\to&(x,y+1),\\
 \mathcal R_6:(x,y)&\to&(x-y,x),\\
 \sigma:(x,y)&\to&(y,x).
\end{subeqnarray}
The algebraic relations in $\cal X$ are:
\begin{subeqnarray}
 \mathcal V_1\mathcal V_2&=&\mathcal V_2\mathcal V_1,\\
 \sigma^2&=&I\\
 \mathcal R_6^6&=&I, \\
 \mathcal V_1\mathcal R_6&=&\mathcal R_6\mathcal V_2^{-1} \slabel{eq:cons-d}\\
 \mathcal V_2\mathcal R_6&=&\mathcal R_6\mathcal V_1\mathcal V_2 \slabel{eq:cons-e}\\
 \mathcal V_1\sigma&=&\sigma \mathcal V_2\\
\mathcal R_6\sigma \mathcal R_6&=&\sigma.
\end{subeqnarray}

Let us now determine the subgroup $\mathcal X_e$ of transformations which are necessarily even.
It evidently includes $\mathcal V_1^2$, $\mathcal V_2^2$ and $\mathcal R_6^2$ (noted $\mathcal R_3$).
But there are more even transformations in this subgroup. Using Eq.~\ref{eq:cons-e} we find $\epsilon_{\mathcal V_2}\epsilon_{\mathcal R_6}=\epsilon_{\mathcal R_6}\epsilon_{\mathcal V_1}\epsilon_{\mathcal V_2}$, so $\epsilon_{\mathcal V_1}=1$.
In the same way, using Eq.~\ref{eq:cons-d}, we get $\epsilon_{\mathcal V_2}=1$.
Thus $\mathcal X_e$ is generated by $\mathcal V_1$, $\mathcal V_2$ and $\mathcal R_3$.
The algebraic relations in $\mathcal X_e$ are
\begin{subeqnarray}
\label{eq:constraints}
 \mathcal V_1\mathcal V_2&=&\mathcal V_2\mathcal V_1,\\
 \mathcal R_3^3&=&I, \\
 \mathcal R_3\mathcal V_1&=&\mathcal V_2\mathcal R_3,\\
 \mathcal R_3&=&\mathcal V_1\mathcal V_2\mathcal R_3\mathcal V_2.
\end{subeqnarray}
As explained in Sec.~\ref{sec:APSG}, each of these relations gives a constraint on the gauge transformations associated to these generators.
The Eqs.\ref{eq:constraints} imply that for any site $i$:
\begin{subeqnarray}
\label{eq:constraints2}
\theta_{\mathcal V_2}(\mathcal V_1^{-1}i)-\theta_{\mathcal V_2}(i)&=&p_1 \pi,\\
\theta_{\mathcal R_3}(i)+\theta_{\mathcal R_3}(\mathcal R_3i)+\theta_{\mathcal R_3}(\mathcal R^2_3i)&=&p_2\pi,\\
\theta_{\mathcal R_3}(i)-\theta_{\mathcal R_3}(\mathcal V_2^{-1}i)-\theta_{\mathcal V_2}(i)&=&p_3\pi,\\
\theta_{\mathcal V_2}(\mathcal V_1^{-1}i)+\theta_{R_3}(\mathcal V_2^{-1}\mathcal V_1^{-1}i)&&\nonumber\\
+ \theta_{\mathcal V_2}(\mathcal V_2\mathcal R_3^2i)-\theta_{\mathcal R_3}(i)&=&p_4\pi,
\end{subeqnarray}
where $p_1$ to $p_4$ can take either the value $0$ or $1$ (the equations are written modulo $2\pi$).
We note $\lbrack x\rbrack$ the integer part of $x$ and $x^*=x-\lbrack x\rbrack$ ($0\leq x^*<1$).
By partially fixing the gauge, we can impose
\begin{subeqnarray}
\label{eq:partial_gauge_fixing}
 \theta_{\mathcal V_1}(x_i,y_i)&=&0
 \nonumber\\
 \theta_{\mathcal V_2}(x_i^*,y_i)&=&p_1 \pi x_i^*.
 \nonumber
\end{subeqnarray}
Through a gauge transformation $G$ of argument $\theta_G$, the $\theta_X$ of a lattice transformation $X$ becomes:
\begin{equation}
\label{eq:gauge_effect}
 \theta_X(i) \to \theta_G(i)+\theta_X(i)-\theta_G(X^{-1}i).
\end{equation}
and the algebraic PSG is transformed in an other element of its equivalence class.
Using the following gauge transformations:
\begin{subeqnarray}
 G_3:(x,y)&\to&\pi x,\nonumber\\
 G_4:(x,y)&\to&\pi y,\nonumber
\end{subeqnarray}
we see that a change of $p_3$ or $p_4$ is a gauge transformation, so we can set them to zero.
Solving the set of equations~\ref{eq:constraints2} leads to:
\begin{subeqnarray}
\label{eq:APSG_tri}
 \theta_{\mathcal V_1}(x,y)&=&0 \\
 \theta_{\mathcal V_2}(x,y)&=&p_1\pi x \\
 \theta_{\mathcal R_3}(x,y)&=&
p_1\pi x\left( y-\frac{x+1}{2}\right)+g_{\mathcal R_3}(x^*, y^*),
\end{subeqnarray}
with a supplementary constraint that can only be treated when the spin lattice is defined:
\begin{eqnarray}
\label{eq:constraintgR}
g_{\mathcal R_3}(x^*, y^*)
+g_{\mathcal R_3}((-y)^*, (x-y)^*)&&\nonumber\\
+g_{\mathcal R_3}((y-x)^*,(-x)^*) &=&p_2\pi.
\end{eqnarray}
This constraint only depends on the coordinates of the sites in a unit cell ($x^*$ and $y^*$).

 Eqs.~\ref{eq:APSG_tri} and \ref{eq:constraintgR} define the chiral algebraic PSG on the triangular Bravais lattice.
 The full determination of the WS Ants\"atze requires  precise definition of the spin lattice (triangular, honeycomb ($m=2$) or kagome ($m=3$)) and on the number of interactions included
 in the MF Hamiltonian (first neighbor only or first and second neighbor; $ \cal{ A}$ and $\cal {B}$ parameters, or $\cal{ A}$ only...).
 The case of the triangular lattice ($m=1$) with nearest neighbor interactions and $\cal{ A}$ and $\cal {B}$ MF parameters is described in the next subsection.

\section{ Strictly and Weakly symmetric Ans\"atze on the triangular lattice with first neighbor interactions}
\label{sec:ansatze_tri}
\subsection{Construction of WS Ans\"atze on the triangular lattice}

The triangular lattice has a single site per unit cell and the values of $x^*$ and $y^*$ are the coordinates of this site in a unit cell, say $(0,0)$.
Eq.~\ref{eq:constraintgR} simplifies into:
\begin{equation}
 6g_{\mathcal R_3}(0,0)=0.
\end{equation}
The solutions are $g_{\mathcal R_3}(0,0)=k\pi/3$, with $k$ integer. Because the IGG is $\mathbb Z_2$, only the three values $k=-1,0,1$ lead to physically different Ans\"atze.

\begin{figure}
\begin{center}
\includegraphics[width=.28\textwidth]{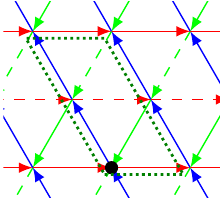}
 \caption{(Color online)
 Ans\"atze respecting the $\mathcal X_e$ symmetries on the triangular lattice.
 All arrows carry $\mathcal B_{ij}$ parameters of modulus $B_1$ and of argument $\phi_{B_1}$ and $\mathcal A_{ij}$ parameters of modulus $A_1$ and of argument 0 on red arrows (choice of the gauge),
 $2k\pi/3$ on blue ones and $4k\pi/3$ on green ones.
 On dashed arrows $\mathcal A_{ij}$ and $\mathcal B_{ij}$ take an extra $p_1\pi$ phase.
 }
\label{fig:triangular_ansatz}
\end{center}
\end{figure}

Finally, we have 6 distinct algebraic PSG's for the reduced set of symmetries ${\cal X}_e$.
They are characterised by two integers $p_1=0,1$ and $k=-1,0,1$ and defined by:
\begin{subeqnarray}
 \theta_{\mathcal V_1}(x,y)&=&0 \\
 \theta_{\mathcal V_2}(x,y)&=&p_1\pi x \\
 \theta_{\mathcal R_3}(x,y)&=&p_1\pi x\left(y-\frac{x+1}{2}\right)+\frac{k\pi}{3}
\end{subeqnarray}
Now, we have to find all the Ans\"atze compatible with these PSG's.
\footnote{In the most general case, an Ansatz contains MF parameters for any couple of sites $(i,j)$, but in practice, we study short range interactions.
For this example, we limit ourselves to first neighbor parameters, but this procedure is easily generalized to further neighbors.}
The first useful insight is to count the number of independent bonds.
Here, one can obtain any bond from any other by a series of rotations and translations ({\it i.e.}, elements of $\mathcal{\chi}_e$).
Thus, if we fix the value of $\mathcal A_{ij}$ and $\mathcal B_{ij}$ on a bond $ij$, we can deduce all other bond parameters from the PSG.
Note that $\mathcal A_{ij}$ can be chosen real by using the gauge freedom.
The value of all bond parameters are represented on Fig.~\ref{fig:triangular_ansatz} as a function of their value on the reference bond.
The  unit cell of the Ansatz contains up to two sites because $p_1$ may be non-zero.

From now on we can forget about the PSG construction and only retain the definition of the Ansatz given by  Fig.~\ref{fig:triangular_ansatz} and
its minimal set of parameters: two integers $p_1$ and $k$, two modulus $A_1$ and $B_1$, and one argument $\phi_{B_1}$.

Until now, we have only considered the subgroup $\mathcal X_e$ and we have looked for Ans\"atze strictly respecting these symmetries.
We now want to consider all symmetries in $\mathcal X$, but the symmetries in $\mathcal X_o$ will be obeyed modulo an eventual time-reversal symmetry.
This requires supplementary conditions on the Ans\"atze of Fig.~\ref{fig:triangular_ansatz}.
As explained in Sec.~\ref{sec:CSL}, the transformations of $\mathcal X_o$ imply relations between the modulus and the arguments of the Ansatz.
Since we are in a very simple case, where all bonds are equivalent in $\mathcal X_e$, no extra relation on the modulus can be extracted from $\mathcal X_o$.
However, some conditions can be found by examining how the the fluxes
$\rm{Arg}(\mathcal A_{ij}\mathcal A_{jk}^*\mathcal A_{kl}\mathcal A_{li}^*)$ on an elementary rhomboedron and $\rm{Arg}(\mathcal A_{ij}\mathcal B_{jk}\mathcal A_{ki}^*)$ on an elementary triangle
transform with $\mathcal R_6$ and $\sigma$. Assuming that neither $A_1$ nor $B_1$ are zero we find:
\begin{subeqnarray}
\label{eq:epsilon}
 2 k\pi(1-\epsilon_{\mathcal R_6})/3&=&0\\
 2 k\pi(1+\epsilon_\sigma)/3&=&0\\
 (1+\epsilon_{\mathcal R_6})\phi_{B1}&=&p_1\pi\\
 (1-\epsilon_\sigma)\phi_{B_1}&=&p_1\pi
\end{subeqnarray}
For each set $(\epsilon_{\mathcal R_6},\epsilon_\sigma)$, the compatible Ans\"atze are thus limited to:\\
i) $(\epsilon_{\mathcal R_6},\epsilon_\sigma)=(1,1)$: $k=0$, $p_1=0$ and $\phi_{B_1}=0$ or $\pi$,\\
ii) $(\epsilon_{\mathcal R_6},\epsilon_\sigma)=(-1,-1)$: $k=0$, $p_1=0$ and $\phi_{B_1}=0$ or $\pi$,\\
iii) $(\epsilon_{\mathcal R_6},\epsilon_\sigma)=(1,-1)$: $\phi_{B_1}=p_1\pi/2$ or $\pi+p_1\pi/2$,\\
iv) $(\epsilon_{\mathcal R_6},\epsilon_\sigma)=(-1,1)$: $k=0$, $p_1=0$ and no constraint on $\phi_{B_1}$.

A couple $(\epsilon_{\mathcal R_6},\epsilon_\sigma)$ does not characterize an Ansatz. A given Ansatz, can be found for several couples of parities.
For example, the Ans\"atze obtained for $(\epsilon_{\mathcal R_6},\epsilon_\sigma)=(1,1)$ are also present for all other $(\epsilon_{\mathcal R_6},\epsilon_\sigma)$.
Indeed as their MF parameters are real, they are not sensitive to time reversal and any $\epsilon_{\mathcal R_6}$, $\epsilon_\sigma$ can be chosen.
From the classical point of view, these Ans\"atze describe coplanar spin configurations, which are invariant under a global spin flip followed by a $\pi$ rotation around an axis perpendicular to the spin plan.

Finally, there are nine different WS Ans\"atze families, given in Table~\ref{tab:symmetric_Ansatze_tri}.
We now conclude this section by a series of remarks concerning the solutions we have obtained:
\begin{itemize}
\item[i)] The number of WS Ans\"atze families is larger than the number of algebraic PSG of $\cal X _\textrm{e }$, because the operators in $\mathcal X_o$ can act in different ways on the Ans\"atze.
\item[ii)] Amongst these 9 Ans\"atze families, only the two first are non chiral, and the 6 others are TRSB Ans\"atze (by applying $\cal T$, $k=1$ is changed to $k=-1$ and $\phi_{B_1}$ to $-\phi_{B_1}$).
The 6 families obtained by aplying $\cal T$ are not listed here.
\item[iii)] These solutions are called {\it families} as the moduli $A_1$ and $B_1$ can vary continuously without modifying the symmetries.
The third Ansatz has no fixed value for $\phi_{B_1}$ and includes the first and second Ans\"atze families (they are kept as distinct as they are non chiral).
\item[iv)] The fluxes of these Ans\"atze are easily calculated using Fig.~\ref{fig:triangular_ansatz}.
\item[v)] The detailed list of compatible Ans\"atze depends on the choice of the mean-field parameters (here, non zero $\mathcal A_{ij}$ and  $\mathcal B_{ij}$ on first neighbor bonds) as we explain in App:~\ref{app:strange} by contrasting these results to those of Wang \textit{et al.} on the same lattice.\cite{PSG}
\end{itemize}

\begin{table}
\renewcommand{\arraystretch}{1.25}
\begin{center}
\begin{tabular}{|c|c|c|c|}
\hline
Ansatz number&~~$p_1$~~ &~~$k$ ~~&~~$\phi_{B_1}$~~ \\
\hline
1&\multirow{5}{*}{0}&\multirow{3}{*}{0}&0\\
\cline{4-4}
2&&&$\pi$\\
\cline{4-4}
3&&&any\\
\cline{3-4}
4&&\multirow{2}{*}{1}&0\\
\cline{4-4}
5&&&$\pi$\\
\cline{2-4}
6&\multirow{4}{*}{1}&\multirow{2}{*}{0}&$\pi/2$\\
\cline{4-4}
7&&&$3\pi/2$\\
\cline{3-4}
8&&\multirow{2}{*}{1}&$\pi/2$\\
\cline{4-4}
9&&&$3\pi/2$\\
\hline
\end{tabular}
\caption{The nine weakly symmetric Ans\"atze families on the triangular lattice, with the notations of Fig.~\ref{fig:triangular_ansatz}. The modulus $A_1$ and $B_1$ are not constrained although supposed non zero.
\label{tab:symmetric_Ansatze_tri}
}
\end{center}
\end{table}

\begin{table}
\renewcommand{\arraystretch}{1.25}
\begin{center}
\begin{tabular}{|c|c|c|c|c|c|c|}
\hline
~~~~~~&~~~F~~~ &coplanar &~~tetra~~\\
\hline
$p_1$ &0   &0   &1\\
$k$&-   &0   &1\\
$\varepsilon_R$   &?   &?   &1\\
$\varepsilon_\sigma$&?   &?   &-1\\
\hline
$A_1$  &0 &$\frac{\sqrt3}{4}^*$ &$\frac1{\sqrt6}^*$\\
\hline
$B_1$  &$\frac12^*$&$\frac14^*$ &$\frac1{\sqrt{12}}^*$\\
$\phi_{B_1}$ &0  &$\pi$  &$\frac\pi2$\\
\hline
\end{tabular}
\caption{Values of the parameters of Fig.~\ref{fig:triangular_ansatz} for Ansatz families related to regular classical states on the triangular lattice.
The states are designed by F for ferromagnetic, coplanar is the $\sqrt3\times\sqrt3$ state and tetra for tetrahedral.
These states are described in more details in~\onlinecite{Regular_order}.
The interrogation points mean that the two values $\epsilon=\pm1$ are possible (coplanar or colinear state).
The $^*$ means that the parameter value is free, we give its value in the fully magnetized state ($|\mathbf m|=1$).
\label{tab:regular_states_tri}
}
\end{center}
\end{table}

\subsection{Condensation of the WS Ans\"atze: the missing tetrahedral state}
\label{sec:tetrahedron}

The SBMFT has already been used to study the antiferromagnetic Heisenberg first-neighbor Hamiltonian on the triangular lattice with the $\mathcal A_{ij}$-only decoupling\cite{Sachdev, PSG}
(Eq.~\ref{eq:decoupling_A}) or with both $\mathcal A_{ij}$ and $\mathcal B_{ij}$\cite{Trumper_AB_SBMFT} (Eq.~\ref{eq:decoupling_AB}).
The classical limit of this model gives the well known three sublattice N\'eel order with coplanar spins at angles of $120^\circ$.
The bond parameters obtained from this classical order (see Eq.~\ref{eq:AB_class}) lead to a strictly symmetric Ansatz
(no need to break $\cal T$: we can chose to fix $(\epsilon_{\mathcal R_6},\epsilon_\sigma)=(1,1)$) with $p_1=0$, $k=0$ and $\phi_{B_1}=\pi$.
We note that all MF parameters are real in this gauge choice (is it always possible to do so for coplanar states).
In this case the restriction to real bond parameters did not prevent to obtain the true MF ground state.

The tetrahedral state (Fig.~\ref{fig:order_tri_tetra})
is the unique GS of the multi-spin exchange Hamiltonian in a large range of parameters:\cite{Momoi_classique,Regular_order}
\begin{equation}
  \widehat H=J_2\sum_{\langle i j \rangle} \hat P_{(ij)} + J_4 \sum_{\langle ijkl\rangle}(\hat P_{(ijkl)}+\hat P_{(ilkj)}),
 \label{eq:Ham_J2J4}
\end{equation}
where the second sum runs on every elementary rhomboedra and $\hat P_{(ijkl)}$ is a cyclic permutation of the spins and $J_4>0$ and $\frac14<\frac{J_2}{J_4}<1$.
Moreover, it is one of the GS's of a Heisenberg Hamiltonian with first and second neighbor interactions
\begin{equation}
 \widehat H=\sum_{\langle i j \rangle}\widehat{\mathbf S}_i\cdot \widehat{\mathbf S}_j+\alpha\sum_{\langle\langle i j \rangle\rangle}\widehat{\mathbf S}_i\cdot \widehat{\mathbf S}_j
\end{equation}
for $\frac18\leq\alpha\leq1$. In the later situation the GS is however degenerate and fluctuations (order by disorder) favors collinear orders.\cite{Jolicoeur_J1J2tri,Lecheminant1995}

The bond parameters obtained from this classical order (Eq.~\ref{eq:AB_class}) lead to the weakly symmetric Ansatz
($(\epsilon_{\mathcal R_6},\epsilon_\sigma)=(1,-1)$) with $p_1=1$, $k=1$ and $\phi_{B_1}=\pi/2$ (or opposite $k$ and $\phi_{B_1}$ for the opposite chirality).
The previous SBMFT studies of the ring exchange model (Eq.~\ref{eq:Ham_J2J4}) have been limited to real parameters\cite{Misguich_SBMFT}
and it would be interesting to perform a  systematic search for a possible chiral MF ground-state.
If the chiral Ansatz indeed turns out to have the lowest energy -- as suggested by its classical limit -- then the spin-$\frac{1}{2}$ might be
a chiral SL since exact diagonalisations\cite{Misguich1999,Misguichmambrini2002} have shown  the absence of Néel long range order in some parameter range.

\section{Fluxes}
\label{sec:fluxes}

We have already given a brief definition of the fluxes in Sec.~\ref{sec:IGG}; in this section we will enlarge this definition and comment on the physical meaning of the various loop operators (local and non local) that can be defined on a lattice.

The gauge invariance of a product of $\widehat A_{ij}$, $\widehat A^\dag_{ij}$, $\widehat B_{ij}$ and $\widehat B^\dag_{ij}$ operators on a closed contour  requires two conditions:
(i) each site $i$ appears in an even number of terms,
(ii) The set of operators containing a site $i$ can be organized into pairs such as the product of each pair is invariant by a local gauge transformation on site $i$ (for example $\widehat A_{ji}$ and $\widehat B_{ik}$).
Such a gauge-invariant operator is the analog of a Wilson loop operator in gauge theory and
the  complex argument of its expectation value is called a flux.
${\rm Arg}\langle \widehat A_{ij}\widehat A^\dag_{jk}\dots\widehat A_{lm}\widehat A^\dag_{mi}\rangle$,
${\rm Arg}\langle \widehat B_{ij}\widehat B_{jk}\dots\widehat B_{li}\rangle$ are examples of fluxes with only $\widehat A_{ij}$ or $\widehat B_{ij}$ operators,
but it is possible to mix both as for example in ${\rm Arg}\langle \widehat A_{ij}\widehat A^\dag_{jk}\widehat B^\dag_{kl}\widehat A_{lm}\widehat A^\dag_{mi}\rangle$.
In SBMFT we approximate these averages of products by the product of the averages (this can be formally justified in the $N\to\infty$ limit).
For example: $\langle \widehat B_{ij}\widehat B_{jk}\dots\widehat B_{li}\rangle \to\mathcal B_{ij} \mathcal B_{jk}\dots \mathcal B_{li}$.

There is an infinite set of non-independent fluxes.\footnote{For example, we can deduce $\rm{Arg}(\mathcal A_{ij}\mathcal B_{jk}\mathcal A^*_{ki}\mathcal A_{im}\mathcal B_{mn}\mathcal A^*_{ni})$
from ${\rm Arg}(\mathcal A_{ij}\mathcal B_{jk}\mathcal A^*_{ki})$ and $\rm{Arg}(\mathcal A_{im}\mathcal B_{mn}\mathcal A^*_{ni})$.
We can thus limit ourselves   to fluxes/loops such as each site is encountered exaclty twice.
Still this is not enough to have  independent fluxes, as the fluxes of two loops having a common part, with the same operators on the common bonds is equal to the flux of the loop encircling both loops.
For example, we can deduce $\rm{Arg}(\mathcal A_{ij}\mathcal A^*_{jk}\mathcal A_{kl}\mathcal A^*_{li})$ from $\rm{Arg}(\mathcal A_{ij}\mathcal A^*_{jk}\mathcal B^*_{ki})$
and $\rm{Arg}(\mathcal A_{kl}\mathcal A^*_{li}\mathcal B^*_{ik})$ (using $\mathcal B_{ik}=\mathcal B^*_{ki}$).}
A method to determine the number of independent fluxes for a given set of non zero $\mathcal A_{ij}$ and $\mathcal B_{ij}$ is given in App.~\ref{app:ind_fluxes}.
To characterize a given Ansatz, we can limit ourselves to the minimal set of independent parameters that define unequivocally its equivalence class: essentially the non-zero bond field modulus and a minimal set of fluxes.

The first insight on the physical meaning of the fluxes is given in the classical limit (Sec.~\ref{sec:flux_class}), where they are simple geometric quantities related to the orientation of the spins.
Then, we come back to the quantum case and express the fluxes, which are physical quantities, with the exclusive use of  spin operators (Sec.~\ref{sec:flux_quantum}).

\subsection{Definition and physical meaning in the classical limit}
\label{sec:flux_class}
We first concentrate on the mean-field flux formed by products of $\mathcal B_{ij}$ parameters.
In the classical limit, the flux of $\mathcal B_{ij}$ around a loop $ijk\dots l$: $\rm{Arg}( \mathcal B_{ij}\mathcal B_{jk}\dots\mathcal B_{li})$
is related to the solid angle associated to the contour described by the spins on the Bloch sphere.
We give here a simplified formulation of the calculation given in Ref.~\onlinecite{Auerbach}.
Let us suppose that the direction of the magnetization (with a modulus fixed to 1) evolves slowly along the loop and use the gauge of Eq.~\ref{eq:ab_class}, but in spherical coordinates:
\begin{equation}
\left(
\begin{array}{c}
\langle \widehat b_{i\uparrow}\rangle\\
\langle \widehat b_{i\downarrow}\rangle
\end{array}\right)=\sqrt{S}
\left(\begin{array}{c}
\cos \frac{\theta_i}{2} \\
\sin \frac{\theta_i}{2}e^{i\phi_i}
\end{array}\right).
\label{eq:ab_class2}
\end{equation}
Then:
\begin{equation}
 \rm{Arg}(\mathcal B^*_{ij})\simeq S(1-\cos\theta_i)\frac{\phi_j-\phi_i}{2}.
\label{eq:elem_solid_angle}
\end{equation}
This last quantity (to first order in the variation of the spin) is the half of the solid angle between the three directions defined by the $z$ axis and the spins at site $i$ and $j$.
By summing such quantities around a closed contour, we obtain the half of the solid angle spanned by the spins along the loop.
This  illustrates the gauge dependence of a single $\mathcal B^*_{ij}$: by a gauge transformation we change the direction of the $z$ axis and thus $\rm{Arg}(\mathcal B^*_{ij})$, but the total
solid angle of the closed loop is independent of the choice of the $z$.

In a similar approach the flux $\rm{Arg}(\mathcal A_{ij}(-\mathcal A^*_{jk})\dots\mathcal A_{lm}(-\mathcal A^*_{mi})$ is associated to the half of the solid angle defined by the spins along the loop,
but after  flipping one spin every two sites
(the $j$ spin for $\mathcal A_{ij}$, the $i$ for $-\mathcal A_{ij}^*$).
The $-1$'s present in the above expression have their importance as they can lead to a final difference of $\pi$.

For more complicated fluxes mixing $\mathcal A_{ij}$ and $\mathcal B_{ij}$ parameters,
we flip one spin every two sites on $\mathcal A_{ij}$ and $\mathcal A_{ij}^*$ bonds (as previously),
we flip all of them for $\mathcal B_{ij}$, and none for $\mathcal B_{ij}^*$. The flux is then half the solid angle associated to these modified spin directions.

We can now  reformulate the previously discussed relation between chirality and fluxes.
If a classical state is chiral, it has non trivial fluxes on contours where the spins are non coplanar.
If the corresponding MF parameters are non zero, we then have found a loop with a non-trivial flux and whatever the gauge choice, at least one MF parameter has to be complex.
Now, if a state is coplanar, then all fluxes are trivial and in a gauge where the spin plane is $xz$, all MF parameters are real.

In the tetrahedral state described on Fig.~\ref{fig:order_tri_tetra}, the flux of the $\mathcal A_{ij}$ around a small rhomboedron is $\pm\pi/3$ and
the flux of the $\mathcal B_{ij}$ around a small triangle is $\pm\pi/2$ (depending on the choice $k=\pm1$, see Sec.~\ref{sec:tetrahedron}).

\subsection{Fluxes in quantum models}
\label{sec:flux_quantum}

In the quantum realm, the fluxes can no longer be expressed in term of solid angles.
But as we have already noted, Wilson loop operators are gauge invariant quantities and as such, they are physical observables and can be expressed in terms of the spin operators.
\subsubsection{Spin-1/2 formulas}
To simplify we will start by imposing that the constraint is strictly verified for $S=\frac{1}{2}$, so there is exactly one boson per site.
We have noted that in the classical limit, the scalar chiralities are associated to the fluxes.
In the quantum case, we can express the flux operators in term of permutation operators, generalizing some results of Ref.~\onlinecite{Wen_Wilczek}.
The operator that transports the spins at sites $1,2,3$ to sites $2,3,1$ is the permutation noted $\widehat P_{(123)}$.
We recall that the permutation operator of spins between two sites can be written as:
\begin{equation}
 \widehat P_{(ij)}=\frac12+2\widehat{\mathbf S}_i\cdot\widehat{\mathbf S}_j
\label{eq:perm}
\end{equation}
This straightforwardly implies that the flux of the $\widehat B_{ij}$ operators is
\begin{equation}
:\widehat B^\dag_{12}\widehat B^\dag_{23}...\widehat B^\dag_{n1}: = \frac{1}{2^{n}}\widehat P_{(12..n)}
\label{eq:perm_B}
\end{equation}
The formula for the flux of the $\widehat A_{ij}$ operators is more involved. It reads
\begin{equation}
\begin{array}{rl}
:\widehat A_{12}^\dag\widehat A_{23} \widehat A_{34}^\dag&\dots\widehat A_{2n\,1}:\,= \frac{1}{2^{2n}}
\widehat P_{(12..2n)}(1-\widehat P_{(23)})
\\
&(1-\widehat P_{(45)}) \dots
(1-\widehat P_{(2n\,1)}).
\end{array}
\label{eq:flux_SU2}
\end{equation}
To prove this last assertion, we first note that $\frac{1-\widehat P_{(ij)}}{2}$ is the projector on the singlet state of the two spins $i$ and $j$.
We then verify this equality in the basis of states $\bigotimes_{i=1}^n \psi_{2i,2i+1}$, where $\psi_{i,j}$ are eigen vectors of $P_{(ij)}$.
In the case where at least one bond is in a symmetric state (triplet), both sides of Eq.~\ref{eq:flux_SU2} are zero.
The final step is simply to check that the relation holds for the state which is a product of singlets.

\subsubsection{Fluxes in quantum spin S models}
\label{sec:flux_S}
For $S>1/2$, Eq.~\ref{eq:perm} is no more valid and Eqs.~\ref{eq:perm_B} and \ref{eq:flux_SU2} are not more valid either.
But we can still replace the on-site number of bosons by $2S$ and obtain an expression depending only on the spin operators.
The expression of the product of four $\widehat A_{ij}$ operators is:
\begin{equation}
\label{eq:Aord}
\begin{array}{c}
8:\widehat A_{12}^\dag\widehat A_{23} \widehat A^\dag_{34}\widehat A_{41}:=
({\mathbf S}_1. {\mathbf S}_2)({\mathbf S}_3. {\mathbf S}_4)
+({\mathbf S}_2. {\mathbf S}_3)({\mathbf S}_4. {\mathbf S}_1)\nonumber\\
-({\mathbf S}_1. {\mathbf S}_3)({\mathbf S}_2. {\mathbf S}_4)
+S^2({\mathbf S}_1 .{\mathbf S}_3+{\mathbf S}_2. {\mathbf S}_4
-{\mathbf S}_1. {\mathbf S}_2
\nonumber \\
-{\mathbf S}_2. {\mathbf S}_3
-{\mathbf S}_3. {\mathbf S}_4
-{\mathbf S}_4. {\mathbf S}_1)
+S^4
+iS(
{\mathbf S}_4.({\mathbf S}_1 \times {\mathbf S}_2)
\nonumber \\
-{\mathbf S}_1.({\mathbf S}_2 \times {\mathbf S}_3)
+{\mathbf S}_2.({\mathbf S}_3 \times {\mathbf S}_4)
-{\mathbf S}_3.({\mathbf S}_4 \times {\mathbf S}_1)
)
\end{array}
\end{equation}
The expression of the product of three $\widehat B_{ij}$ operators is:
\begin{equation}
\label{eq:Bord}
\begin{array}{c}
4:\widehat B^\dag_{12}\widehat B^\dag_{23}\widehat B^\dag_{31}:=
S({\mathbf S}_1 .{\mathbf S}_2+{\mathbf S}_2. {\mathbf S}_3+{\mathbf S}_3. {\mathbf S}_1)
\\
+S^3
-i{\mathbf S}_1.({\mathbf S}_2 \times {\mathbf S}_3)
\end{array}
\end{equation}

\subsubsection{Fluxes in SBMFT}
\label{sec:flux_SBMFT}
In a state where on-site number of bosons is not strictly conserved, the previous expressions become a bit more complicated.
The number operators can no longer be replaced by $2S$, and we have for example:
\begin{equation}
\label{eq:Bord2}
\begin{array}{c}
4:\widehat B^\dag_{12}\widehat B^\dag_{23}\widehat B^\dag_{31}:=
\frac12 \widehat n_3{\mathbf S}_1 .{\mathbf S}_2+\frac12 \widehat n_1{\mathbf S}_2. {\mathbf S}_3+\frac12 \widehat n_2{\mathbf S}_3. {\mathbf S}_1
\\
+\frac{\widehat n_1\widehat n_2\widehat n_3}8
-i{\mathbf S}_1.({\mathbf S}_2 \times {\mathbf S}_3).
\end{array}
\end{equation}

\subsection{Finite size calculations lattice symmetries and non local fluxes}
\label{sec:flux_periodicity}

For simple lattices as the square or triangular lattice, we can solve analytically the MF Hamiltonian $H_{\rm{MF}}$ of Eq.~\ref{eq:HMF} directly in the thermodynamical limit.
But in most cases, we have to solve numerically the self-consistency conditions on finite lattices.

To use the chiral PSG's on a finite periodic lattice, we have to be  cautious about symmetries.
Indeed, all  precautions have been taken so that the Ansatz (strictly or weakly) respects the lattice symmetries on an infinite lattice.
But we have to verify that the finite periodic lattice has the same symmetry group as the infinite one.
This verification is quite usual for local properties, but is more subtle for non-local ones and can be most easily understood in term of fluxes on large non-local loops.

PSG's impose  that fluxes on local loops are preserved by lattice symmetries (or sent to their opposite in the case of a chiral state).
But some additional care has been taken concerning loops
which are topologically non trivial (cannot be shrunk to a point by a succession of local deformations).
These loops which ``winds'' through the  boundary conditions  do not exist on the infinite lattice.
For a symmetric Ansatz to remain symmetric on a finite periodic lattice, we  have to verify that the fluxes associated to these topologically non trivial loops also respect the lattice symmetries.
The way to treat the problem of the non-local loops is detailed in App.~\ref{app:fluxes}, together with several ways of understanding their meaning.

\section{Conclusion}
\label{sec:ccl}

In this paper we have extended the PSG construction to include  time-reversal-symmetry-breaking states with the SBMFT.
These TRSB phases that we describe generically as \textit{chiral }, can also break one or many discrete symmetries of the lattice
(in the triangular example either $\sigma$ or $\cal R_\textrm{6}$). Using this constructive method we have built all the SS and WS Ans\"atze with two MF parameters on the triangular lattice.
All the regular $O(3)$ magnetically ordered phases can be obtained from these Ans\"atze by spinon-condensation  (the others have no regular classical limit).
The TRSB Ans\"atze have, when they condense, non-planar magnetic order and non-zero scalar chiralities.

The TRSB SL have  short range spin-spin correlations but non trivial fluxes on various loops. The simplest of these fluxes are related to the imaginary part of the
permutation operator of three spins, that is directly related to their scalar chirality.
In some cases the time-reversal symmetry breaking fluxes might be more complex, as explained in section \ref{sec:fluxes} and illustrated in Appendix \ref{app:kagome_ansatz} for the kagome lattice.
These various fluxes have been initially defined within the MF Schwinger boson approach but \ref{sec:fluxes} has shown how  these gauge invariant
quantities can be expressed in terms of spin operators, independently of any MF approximation.
It should be noticed that in a TRSB SL fluxes other than those deduced from the Ansatz may be non zero and easier to compute. It is the case for example in the \cuboc SL recently proposed for the nearest-neighbor Heisenberg model on the kagome lattice.\cite{cuboc1}
The flux of the $\widehat A$ bond operators around the hexagons can be expressed in terms of spin permutation operators but it is relatively involved
(Eq.~\ref{eq:flux_SU2}) and has not yet been computed numerically.
In fact, in that phase (at least at the MF level), there are simpler fluxes which are non zero, as for example the triple product of 2nd neighbor spins around hexagons, or the triple product of three consecutive spins on an hexagon.

In Sec.~\ref{sec:tetrahedron}, another  TRSB Ansatz was discussed in relation to the ring exchange model on the triangular lattice.

In spite of short range spin-spin correlations the TRSB SL have some local order parameter associated to  the fluxes.
The finite temperature broken symmetries being discrete symmetries, there are no  Goldstone modes and these chiral phases should survive thermal fluctuations in 2D.
The phase transition associated to the restoration of the  chiral symmetry has been studied in some classical spin models.\cite{KagomeDomenge,MessioDomenge,Triedres}
In spite of the Ising like character of the order parameter, the phase transition was shown to be weekly first order due to interplay of vortices in the magnetic texture
with domain walls of the chirality. It has been shown within the SBMFT framework in the \cuboc
phase that thermal fluctuations tend to do expel the chiral fluxes\cite{cuboc1} (favour coplanar correlations)
but a more complete  study (beyond MF) of the finite temperature properties of a TRSB SL would be required to understand the
specific properties of the chiral transition in these systems.

Finally, it would be useful to clarify the ``topological'' differences (entanglement, degeneracy, edge modes, ...) between the present chiral SL described in the SBMFT framework with
the  chiral SL wave-functions related to fractional quantum Hall states (such as the Kalmeyer-Laughlin state\cite{Kalmeyer_89} or that of Yang, Warman and Girvin\cite{Yang1993} for instances),
as well as the difference with conventional ($\mathcal T$-symmetric) $\mathbb{Z}_2$ liquids.
It would also be very interesting to analyze qualitatively the  effects of (gauge) fluctuations in the present chiral SL.

\appendix

\section{The Bogoliubov transformation}
\label{app:bogoliubov}

This appendix explains how to obtain the eigenmodes of Eq.~\ref{eq:HMF}.
New bosonic operators, components of $\tilde\phi$, are created by linear combinations of the components of $\phi$ to obtain a new diagonal matrix $\tilde M$.
For the Hamiltonian to possess a GS (spectrum bounded from below), the diagonal elements $(\omega_1,\dots \omega_{2N_s})$ must all be positive or null.
This transformation is called the Bogoliubov transformation and is generally well documented (see for instance Ref.~\onlinecite{Auerbach}) when the size of the matrix $M$ is $2\times 2$
(the transformation can then be done analytically), but more rarely for larger sizes (where numerical calculations are sometimes required).
When periodic Ans\"atze are considered, a Fourier transform can block-diagonalize $M$, with blocks of size $2m\times2m$, with $m$ the number of sites in the unit cell.
As soon as $m>1$, Bogoliubov transformation of matrices larger than $2\times2$ are needed.

Note that the choice of an Ansatz whithout any $\mathcal A_{ij}$ parameters
(for example using Eq.~\ref{eq:decoupling_B}) simplifies considerably the Bogoliubov transformation
since the total number of boson is conserved and $M$ is block diagonal with two blocks of size $N_s$.
The transformation reduces to the diagonalization of each block by a unitary  matrix.
The new bosons $\tilde b_{i\sigma}$ are then linear combinations of the old $b_{i\sigma}$, without any $b^\dag_{i\sigma}$ component.
The vacuum of the new bosons is the same vacuum as for the old bosons.
To respect the constraint on the boson number, we have to create a Bose condensate (see Sec.\ref{sec:classical_order}), which implies long-range magnetic order.
This proves that the $\mathcal A_{ij}$ parameters are necessary to obtain SL.

Here we describe the general method for the cases where $\tilde M$ can have an arbitrary size, as explained in details in by Colpa.\cite{Colpa}
The $2N_s\times 2N_s$ matrix $P$ defined such that $\phi=P\tilde\phi$ is called the transformation matrix.
Let us look at the conditions $P$ should satisfy.
The most evident is that $\tilde M$ must be diagonal, which gives a first constraint.
The second one is that the $N_s$ first components of $\tilde\phi$ must be annihilation operators and the last $N_s$, creation operators.
This gives a constraint on their commutation relations.
The two resulting conditions are
\begin{equation}
 \left\{\begin{array}{l}
 P^\dag M P=\tilde M\\
 P^\dag J P=J
\end{array}
\right.
\end{equation}
where $J$ is the $2N_s\times 2N_s$ diagonal matrix with coefficients $-1$ for the $N_s$ first terms and $1$ for the last $N_s$ elements ($J_{ij}=\lbrack\phi^\dag_i,\phi_j\rbrack$).
The second constraint makes the Bogoliubov transformation different from a diagonalization (where $J$ would be the identity matrix).
It is sometimes called a para-diagonalization.

Here we just recall the main steps of the algorithm\cite{Colpa} to solve these equations:
\begin{itemize}
 \item Verify that $M$ is definite positive. It ensures that the GS is unique (in some cases where $M$ has zero eigenvalues, the GS exists but is not the unique).
 \item Find a complex upper-triangular square matrix $K$ such as $M=K^\dag K$ (Cholesky decomposition of $M$).
 \item Find a unitary matrix $U$ such as $L=U^\dag KJK^\dag U$ is diagonal with it first $N_s$ coefficients positive and the other negative (usual diagonalization of a hermitian matrix).
 \item The solution is $\tilde M=JL$ and $P=K^{-1}U\tilde M^{1/2}$.
\end{itemize}

Using the rotational invariance, we deduce that the $N_s$ first coefficients $(\omega_1,\dots,\omega_{N_s})$ of $\tilde M$ are the same as the $N_s$ last (maybe differently ordered).
The energy of the MF Hamiltonian GS writes:
\begin{equation}
 E_0=\frac12\sum_{i=1}^{N_s}\omega_i+\epsilon_0.
\end{equation}
and its elementary excitations are free bosonic spinons with energies $(\omega_1,\dots,\omega_{N_s})$ and spin $1/2$, from which we can get the free energy at any temperature.
We are now able to look for solutions of Eq.~\ref{eq:self_cons2} and \ref{eq:constraint3}, \textit{i.e.} the stationary points of the free energy with respect to the MF parameters and
with respect to the Lagrange multipliers.

\section{Bounds on self-consistent values of the MF parameters in SBMFT}
\label{app:borne}

The moduli $|\mathcal A_{ij}|$ and  $|\mathcal B_{ij}|$ are  \textit{a priori} unconstrained real numbers in
SBMFT. We  prove here that in a self-consistent Ans\"atze, their moduli cannot exceed an upped bound: $|\mathcal A_{ij}|\leq\frac{\kappa+1}2$ and $|\mathcal B_{ij}|\leq\frac\kappa2$.
These inequalities considerably restrict the domain to explore and facilitate the numerical search for solutions.

Let $|\phi\rangle$ be any normalised bosonic state. We denote by $\langle \widehat O\rangle$ the the expectation value of an operator in this state.
Whatever the operators $\widehat u$ and $\widehat v$ we have
\begin{equation}
|\langle\widehat u\widehat v\rangle|\leq
\frac{\langle\widehat u\widehat u^\dag\rangle+\langle\widehat v^\dag\widehat v\rangle}{2}.
\end{equation}
Applying it to $\widehat A_{ij}$ and $\widehat B_{ij}$, we obtain
\begin{equation}
\label{eq:temp}
|\langle\widehat A_{ij}\rangle|\leq
\frac{\langle\widehat n_{i}+\widehat n_{j}+2 \rangle}{4},\quad
|\langle\widehat B_{ij}\rangle|\leq
\frac{\langle\widehat n_{i}+\widehat n_{j}\rangle}{4}.
\end{equation}
We now take $|\phi\rangle$ as the GS of $H_{\rm MF}$ (Eq.~\ref{eq:HMF}) for some Ansatz.
If the chemical potential is adjusted, $\langle\widehat n_{i}\rangle=\kappa$ on every lattice site.
In the case of self consistent parameters, $|\mathcal A_{ij}|=|\langle\widehat A_{ij}\rangle|$, $|\mathcal B_{ij}|=|\langle\widehat B_{ij}\rangle|$ and Eq.~\ref{eq:temp} leads to
\begin{equation}
|\mathcal A_{ij}|\leq
\frac{\kappa+1}{2},\quad
|\mathcal B_{ij}|\leq
\frac{\kappa}{2}.
\end{equation}

\section{The strange classical limit of the \texorpdfstring{$\pi$}{} flux Ansatz of \textcite{PSG}}
\label{app:strange}

\textcite{PSG} explored all the strictly symmetric Ans\"atze ($(\epsilon_{\mathcal R_6},\epsilon_\sigma)=(1,1)$) with the $\mathcal A_{ij}$
decoupling for first neighbor Heisenberg interactions.
They found two Ans\"atze. The first one is  characterized by a flux $\rm{Arg}(\mathcal A_{ij}\mathcal A_{jk}^*\mathcal A_{kl}\mathcal A_{li}^*)=0$ around a rhomboedra for ($p_1=0$),
giving the 3 sublattice N\'eel order in the classical limit. The second one has a flux $\pi$ rhomboedra ($p_1=1$).
The $\mathcal A_{ij}$ parameters they used are those obtained from Fig.~\ref{fig:triangular_ansatz} with the corresponding value of $p_1$ and $k=0$.

Comparing this to our result for the SS Ans\"atze, we may wonder why do they obtained {\it two} possibilities for $p_1$ (0 or 1) whereas
we found that $p_1=0$  was the only solution for $(\epsilon_{\mathcal R_6},\epsilon_\sigma)=(1,1)$.
The difference comes from the the absence of $\mathcal B_{ij}$ parameter in their MF approach.
The complex phase of $\mathcal B_{ij}$ is then ill defined and only the first of our two constraints (Eq.~\ref{eq:epsilon}) remains.
They thus impose $k=0$, but nothing on $p_1$.
In fact, as this situation is the limit $B_1\to0$ of none of the SS cases we have explored in Sec.~\ref{sec:ansatze_tri},
it appears that the $\pi$-flux  Ansatz is unstable with respect to the introduction of $\mathcal B_{ij}$. In other words, any non-zero value of $\mathcal B_{ij}$
will break at least one lattice symmetry.
The two WS Ans\"atze described in the 6th and 7th lines of Tab.~\ref{tab:symmetric_Ansatze_tri}
corresdpond to this limit.

The nature of the spinon condensation in the $\pi$-flux  Ansatz   could not be completely clarified in  Ref.~\onlinecite{PSG}.
Our understanding is that it is not consistent to impose $\mathcal B_{ij}=0$ to describe ordered states
on a frustrated lattice. The only way to have $|\mathcal B_{ij}|^2=0$ classically is indeed to have anti parallel
spins on all bonds, which is not possible on the triangular lattice.

\section{Weakly symmetric Ans\"atze on some usual lattices}
\label{app:EAPSG}

\subsection{Lattices with a square Bravais lattices}
\label{app:square}

The first step is to find all chiral algebraic PSG's.
We choose the most general case IGG$\sim\mathbb Z_2$ and we suppose that $\widehat H_0$ respects all the lattice symmetries whose generators are described in Fig.~\ref{fig:sym_latt} (right).
The coordinates $(x,y)$ of a point are given in the basis of the the translation vectors $\mathcal V_1$, $\mathcal V_2$ and the action of the generators on the coordinates are
\begin{subeqnarray}
 \mathcal V_1:(x,y)&\to&(x+1,y),\\
 \mathcal V_2:(x,y)&\to&(x,y+1),\\
 \mathcal R_4:(x,y)&\to&(-y,x),\\
 \sigma:(x,y)&\to&(y,x).
\end{subeqnarray}
The algebraic relations between them are
\begin{subeqnarray}
\label{eq:constraints_sq_all}
 \mathcal V_1 \mathcal V_2&=& \mathcal V_2 \mathcal V_1\\
 \mathcal V_2 \mathcal R_4&=& \mathcal R_4 \mathcal V_1\\
 \mathcal R^4_4&=&I\\
 \mathcal V_1 \mathcal R_4 \mathcal V_2&=& \mathcal R_4 \\
 \mathcal V_1\sigma&=&\sigma \mathcal V_2\\
 \mathcal R_4\sigma \mathcal R_4&=&\sigma\\
 \sigma^2&=&I
\end{subeqnarray}

To our knowledge, even the non-chiral algebraic PSG's have not been derived previously.
Here, we directly derive the chiral ones.
From Eq.~\ref{eq:constraints_sq_all}, we deduce that the reduced set of symmetries $\mathcal X_e$ is generated by $\mathcal V_1^2$, $\mathcal V_2^2$ and $\mathcal R_4^2$ (noted $\mathcal V_1'$, $\mathcal V_2'$ and $\mathcal R_2$). Moreover, we find that $\epsilon_{\mathcal V_1}=\epsilon_{\mathcal V_2}$. An Ansatz is characterised by the parities $(\epsilon_{\mathcal V_1}, \epsilon_{\mathcal R}, \epsilon_{\sigma} )$.

The algebraic relations between these generators are
\begin{subeqnarray}
\label{eq:constraints_sq}
 \mathcal V_1'\mathcal V_2'&=&\mathcal V_2'\mathcal V_1',\\
 \mathcal R_2^2&=&I, \\
 \mathcal V_1'\mathcal R_2\mathcal V_1'&=&\mathcal R_2,\\
 \mathcal V_2'\mathcal R_2\mathcal V_2'&=&\mathcal R_2.
\end{subeqnarray}
As explained in Sec.~\ref{sec:APSG}, each of these relations gives a constraint on the gauge transformations associated to the generators.
The constraint from Eq.~\ref{eq:constraints_sq} are then,  for all $i$:
\begin{subeqnarray}
\label{eq:constraints2_sq}
\theta_{\mathcal V_2'}(\mathcal V_1'^{-1}i)-\theta_{\mathcal V_2'}(i)&=&p_1 \pi,\\
\theta_{\mathcal R_2}(i)+\theta_{\mathcal R_2}(\mathcal R_2i)&=&p_2 \pi,\\
\theta_{\mathcal R_2}(\mathcal V_1'^{-1}i)-\theta_{\mathcal R_2}(i)&=&p_3\pi,\\
\theta_{\mathcal V_2'}(\mathcal V_2'i)+\theta_{\mathcal V_2'}(\mathcal R_2i)+ \theta_{\mathcal R_2}(i)-\theta_{\mathcal R_2}(i)&=&p_4\pi.
\end{subeqnarray}
where $p_1,\cdots,p_4$ can take either the value $0$ or $1$ (the equations are written modulo $2\pi$).
We note $\lbrack x\rbrack$ the integer part of $x/2$ and $x^*=x-2\lbrack x\rbrack$ ($0\leq x^*<2$).
By partially fixing the gauge, we impose
\begin{subeqnarray}
 \theta_{\mathcal V_1'}(x_i,y_i)&=&0,
 \nonumber\\
 \theta_{\mathcal V_2'}(x_i^*,y_i)&=&p_1\pi x_i^*.
 \nonumber
\end{subeqnarray}
Contrarily to the triangular lattice, no gauge transformation can here be used to get rid of some $p_i$.

Solving the previous equations~\ref{eq:constraints2_sq} leads us to
\begin{subeqnarray}
 \theta_{\mathcal V_1}(x,y)&=&0, \\
 \theta_{\mathcal V_2}(x,y)&=&p_1\pi x, \\
 \theta_{\mathcal R_2}(x,y)&=&p_3\pi x+p_4\pi y+g_{\mathcal R_2}(x^*,y^*).
\end{subeqnarray}
with a complicated supplementary constraint that can be treated only when the lattice is more precisely defined:
\begin{equation}
\label{eq:constraintgR_sq}
g_{\mathcal R_2}(x^*,y^*)
+g_{\mathcal R_2}((-x)^*, (-y)^*)
=p_2\pi.
\end{equation}
This constraint only depends on the coordinates of the sites in a $2\times2$ unit cell ($x^*$ and $y^*$), so it gives at most $4m$ independent constraints.

These general algebraic PSG's can then be used to find the weakly symmetric Ans\"atze on any lattice with a square Bravais lattice (for example: the square, the Shastry-Sutherland lattice,\dots).

\subsection{Weakly symmetric Ans\"atze on the kagome lattice}
\label{app:kagome_ansatz}

The Bravais lattice of the kagome lattice is triangular, so we use the algebreaic PSG's determined in Sec.~\ref{sec:APSG_tri}.
The unit cell contains three sites.
We choose to place the origin of the frame at the center of an hexagon and the coordinates of the sites in a unit cell are $(\frac12,0)$, $(0,\frac12)$ and $(\frac12,\frac12)$.

Since the sites have non-integer coordinates, it is convenient to transform the Eqs.~\ref{eq:APSG_tri} using the following
gauge transformation (see Eq.~\ref{eq:gauge_effect}):
\begin{equation}
 G_1:(x,y)\to-p_1\pi y x^*.
\end{equation}
The new algebraic PSG is
\begin{subeqnarray}
\label{eq:APSG_kag}
 \theta_{\mathcal V_1}(x,y)&=&0 \\
 \theta_{\mathcal V_2}(x,y)&=&p_1\pi \lbrack x\rbrack \\
 \theta_{\mathcal R_3}(x,y)&=&
p_1\pi \lbrack x\rbrack\left(\lbrack y\rbrack-\frac{\lbrack x\rbrack +1}{2}+[y^*-x^*]\right)\nonumber\\
&&+g_{\mathcal R_3}(x^*, y^*),
\end{subeqnarray}
Even if it seems more complicate than Eq.~\ref{eq:APSG_tri}, it avoids some $p_1\pi/2$ and simplify the future Ans\"atze.
This gauge transformation is equivalent to a different initial choice of $\theta_{\mathcal V_2}(x_i^*,y_i)$ in Eq.~\ref{eq:partial_gauge_fixing}:
\begin{equation}
\label{eq:better_gauge}
 \theta_{\mathcal V_2}(x_i^*,y_i)=0.
\end{equation}
Under the effect of $G$, Eq.~\ref{eq:constraintgR} is modified and gives the constraint
\begin{eqnarray}
\label{eq:constraintgR_kag}
g_{\mathcal R_3}\left(\frac12,0\right)
+g_{\mathcal R_3}\left(0, \frac12\right)
+g_{\mathcal R_3}\left(\frac12,\frac12\right) &=&(p_2+p_1)\pi.\nonumber
\end{eqnarray}
Using the following gauge transformations,
\begin{subeqnarray}
G_2:(x,y)&\to&ax^*,\\
G_3:(x,y)&\to&by^*,\\
G_4:(x,y)&\to&\lbrack y^*-x^*\rbrack \pi,
\end{subeqnarray}
with $a$ and $b$ real numbers, we can set $g_{\mathcal R_3}=0$.
Finally, we have two distinct algebraic PSG's for the reduced set of symmetries.
They are characterised by $p_1=\pm1$ and defined by Eq.~\ref{eq:APSG_kag} with $g_{\mathcal R_3}=0$.

\begin{figure}
\begin{center}
\includegraphics[width=.28\textwidth]{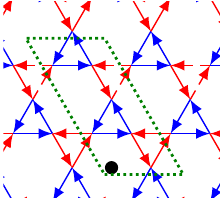}
 \caption{(Color online)
 Ans\"atze respecting the $\mathcal X_e$ symmetries on the kagome lattice. 
 Blue arrows carry $\mathcal B_{ij}$ parameters of modulus $B_1$ and of argument $\phi_{B_1}$ and $\mathcal A_{ij}$ parameters of modulus $A_1$ and of argument $0$.
 Red arrows are for modulus $B_1'$ and $A_1'$ and arguments $\phi_{B_1'}$ and $\phi_{A_1'}$. On dashed arrows $\mathcal A_{ij}$ and $\mathcal B_{ij}$ take an extra $p_1\pi$ phase.
 }
 \label{fig:kagome_ansatz}
\end{center}
\end{figure}

Now, we have to find all Ans\"atze compatible with these PSG's.
We limit ourselves to first neighbor parameters, but this procedure is easily generalized to further neighbors.
Two bonds are needed to generate the whole lattice by action the reduced symmetries: one blue bond and one of red bond  of Fig.~\ref{fig:kagome_ansatz}.
The values of $\mathcal A_{ij}$ and $\mathcal B_{ij}$ on these two reference bonds allows to obtain all the other bond parameters by using the PSG.
Note that $\mathcal A_{ij}$ can be chosen real for say the reference blue bond by using the gauge freedom.
The values of all bond parameters are represented on Fig.~\ref{fig:kagome_ansatz} as a function of their value on the reference bond.
The general unit cell of the parameters contains six sites because of the possibly non-zero $p_1$.
Note that the simplicity of the Ansatz of Fig.~\ref{fig:kagome_ansatz}, where phases differ only by $\pi$
between two bonds of the same color is a consequence of the choice of Eq.~\ref{eq:better_gauge}.

Finally, we can forget all about the PSG and only retain the parameters needed to completely describe an Ansatz together with Fig.~\ref{fig:kagome_ansatz}.
These parameters consist in the integer $p_1$, the modulus $A_1$, $A_1'$, $B_1$ and $B_1'$, and the arguments $\phi_{A_1'}$, $\phi_{B_1}$ and $\phi_{B_1'}$.

We now want to consider all symmetries in $\mathcal X_o$.
As blue and red bonds are related through $\mathcal R_6$, this implies relations between the modulus: $A_1=A_1'$ and $B_1=B_1'$.
Relations on the phases are found by looking at the effect of $\mathcal R_6$ and $\sigma$ on the flux
$\rm{Arg}(\mathcal A_{ij}\mathcal A_{jk}^*\mathcal A_{kl}\mathcal A_{lm}^*\mathcal A_{mn}\mathcal A_{ni}^*)$
of an elementary bow tie and $\rm{Arg}(\mathcal A_{ij}\mathcal B_{jk}\mathcal A_{ki}^*)$ of an elementary triangle (we suppose that neither $A_1$ nor $B_1$ are zero):
\begin{subeqnarray}
(\epsilon_{\mathcal R_6}+\epsilon_\sigma)\phi_{B_1}=0\\
\phi_{B_1'}=\epsilon_{\mathcal R_6}\phi_{B_1}\\
(1+\epsilon_{\mathcal R_6})\phi_{A_1'}=0\\
(1+\epsilon_\sigma)\phi_{A_1'}=0
\end{subeqnarray}
For each couple $(\epsilon_{\mathcal R_6},\epsilon_\sigma)$, the number of compatible Ans\"atze is thus reduced:\\
i) $(\epsilon_{\mathcal R_6},\epsilon_\sigma)=(1,1)$: $\phi_{B_1}=\phi_{B_1}'=0$ or $\pi$ and $\phi_{A_1'}=0$ or $\pi$,\\
ii) $(\epsilon_{\mathcal R_6},\epsilon_\sigma)=(1,-1)$: $\phi_{B_1}=\phi_{B_1}'$ and $\phi_{A_1'}=0$ or $\pi$,\\
iii) $(\epsilon_{\mathcal R_6},\epsilon_\sigma)=(-1,1)$: $\phi_{B_1}=-\phi_{B_1}'$ and $\phi_{A_1'}=0$ or $\pi$,\\
iv) $(\epsilon_{\mathcal R_6},\epsilon_\sigma)=(-1,-1)$: $\phi_{B_1}=\phi_{B_1}'$ and $\phi_{B_1}=0$ or $\pi$.\\

\begin{table}
\renewcommand{\arraystretch}{1.05}
\begin{center}
\begin{tabular}{|c|c|c|c|c|}
\hline
Ansatz number&~~$p_1$~~ &~~$\phi_{A_1}'$~~ &~~$\phi_{B_1}$ ~~&~~$\phi_{B_1}'$~~\\
\hline
1&\multirow{10}{*}{0}&\multirow{4}{*}{0}&\multirow{3}{*}{$\phi_{B_1}'$}&0\\
\cline{5-5}
2&&&&$\pi$\\
\cline{5-5}
3&&&&any\\
\cline{4-5}
4&&&$-\phi_{B_1}'$&any\\
\cline{3-5}
5&&\multirow{4}{*}{$\pi$}&\multirow{3}{*}{$\phi_{B_1}'$}&0\\
\cline{5-5}
6&&&&$\pi$\\
\cline{5-5}
7&&&&any\\
\cline{4-5}
8&&&$-\phi_{B_1}'$&any\\
\cline{3-5}
9&&\multirow{2}{*}{any}&\multirow{2}{*}{$\phi_{B_1}'$}&0\\
\cline{5-5}
10&&&&$\pi$\\
\cline{2-5}
11&&\multicolumn{3}{|c|}{}\\
$\vdots$&1&\multicolumn{3}{|c|}{same as for $p_1=0$}\\
20&&\multicolumn{3}{|c|}{}\\
\hline
\end{tabular}
\caption{The twenty weakly symmetric Ans\"atze families on the kagome lattice, with the notations of Fig.~\ref{fig:kagome_ansatz}.
The modulus $A_1=A_1'$ and $B_1=B_1'$ are not constrained, except that they do not vanish.
\label{tab:symmetric_Ansatze_kag}
}
\end{center}
\end{table}

Finally, there are 20 different WS Ans\"atze families, given in Table~\ref{tab:symmetric_Ansatze_kag}.
Each regular states of Ref.~\onlinecite{Regular_order} belongs to one of them:
the 2nd for the \qzero, the 6th for the \sqrtsqrt, the 17th for the \textit{octahedral}, the 20th for the \cuboc and the 14th for the \cubocd state.
The parameters of the fully magnetized states are calculated using Eq.~\ref{eq:ab_class}.
They are also used to determine the parities and $p_1$.
In the case where both parities are possible (coplanar states), we chose the largest symmetries ($\epsilon=1$) to fix the maximum number of parameters.
They are indicated in Table \ref{tab:regular_states}.
For the parameters which are not fixed, we give the values take in the classical limit.
The self consistent parameters for finite $S$ are different, but generally not far from the classical values.
Thus, they can be used as a starting point in numerical optimizations.

These calculations are easily generalised to further neighbors and have already been used for two studies on the kagome lattice\cite{Faak2012,cuboc1}

\begin{table}
\renewcommand{\arraystretch}{1.25}
\begin{center}
 \begin{tabular}{|c|c|c|c|c|c|c|}
\hline
         &~F~    &$q=0$      &$\sqrt3\sqrt3$&oct       &cuboc1     &cuboc2  \\
\hline
$p_1$       &0    &0        &0&1        &1        &1        \\
$\varepsilon_R$  &?    &?        &?&1        &-1        &-1        \\
$\varepsilon_\sigma$&?   &?        &?&-1        &-1        &1        \\
\hline
$A_1$       &0    &$\frac{\sqrt3}2^*$&$\frac{\sqrt3}2^*$&$\frac1{\sqrt2}^*$&$\frac{\sqrt3}2^*$ &$\frac12^*$ \\
$\phi_{A_1'}$   &$-$   &0        &$\pi$&$\pi$      &$\pi-\rm{atan}{\sqrt8}^*$&0     \\
\hline
$B_1$       &$1^*$  &$\frac12^*$  &$\frac12^*$&$\frac1{\sqrt2}^*$&$\frac12^*$    &$\frac{\sqrt3}2^*$ \\
$\phi_{B_1}$   &0    &$\pi$     &$\pi$&$\frac{-3\pi}4^*$        &$\pi$ &$\rm{atan}{\sqrt2}-\pi^*$\\
$\phi_{B_1'}$   &0    &$\pi$     &$\pi$ &$\phi_{B_1}$ &$\pi$ &$-\phi_{B_1}$   \\
\hline
 \end{tabular}
\caption{Values of the parameters of Fig.~\ref{fig:kagome_ansatz} for Ansatz families related to regular classical states on the kagome lattice (the states are designed by F for ferromagnetic, oct for octahedral and cuboc for cuboctaedron order parameters.
These states are described in more details in \cite{Regular_order}).
The modulus verify $A_1=A_1'$ and $B_1=B_1'$.
The interrogation points mean that the two values $\epsilon=\pm1$ are possible (coplanar state).
The $^*$ mean that the parameter value is free, we give its value in the fully magnetized state ($|\mathbf m|=1$).
\label{tab:regular_states}
}
\end{center}
\end{table}

\section{Number of independent fluxes on a lattice}
\label{app:ind_fluxes}

We suppose that we have a MF Hamiltonian with $n_\mathcal A+n_\mathcal B$ non-zero bond parameters $(\{\mathcal A_{ij}\}, \{\mathcal B_{ij}\})$ ($\mathcal A_{ij}$ and $\mathcal A_{ji}$ count as only one parameter, and the same for $\mathcal B_{ij}$ and $\mathcal B_{ji}$).
As they are complex numbers, we need $2n_\mathcal A+2n_\mathcal B$ self-consistent conditions to solve this MF problem.
We already know that the solution is not unique and that two Ans\"atze related by a gauge transformation are equivalent.
Thus, by fixing the gauge, we can decrease the number of equations for the complexe phases.
In fact, the number $f$ of necessary arguments corresponds to the number of independent fluxes on the lattice.
In this appendix, we describe a simple method to compute $f$ on a finite cluster.

We define a rectangular matrix $M$ of size $(n_\mathcal A+n_\mathcal B)\times N_s$ ($N_s$ is the number of sites), where each line characterizes a MF parameter, and is therefore
associated to a pair of sites $(ij)$.
As for the column, they correspond to the lattice sites.
The coefficients of a line are all zero except for the two entries at columns $i$ and $j$.
Both entries  are equal $1$ for an $\mathcal A_{ji}$-bond, whereas  these entries are $-1$ and $1$ for a $\mathcal B_{ji}$-bond (which site is $\pm$ 1 has no importance).
Then the result is:
\begin{equation}
 f=n_\mathcal A+n_\mathcal B-\rm{rank}(M).
 \label{eq:nb_fluxes}
\end{equation}

The effect of a gauge transformation on the bond phases is obtained by multiplying $M$ by the vector $(\theta_1,\cdots \theta_{N_s})^t$.
By definition, a product of bond parameters defines a flux if the sum of their complex phases is unchanged by a gauge transformation.
It means that the sum of associated matrix lines is 0.
As the complex conjugate of a bond parameter can be used, the weight of each line in the sum can be $\pm1$.
As we can imagine using several times the same parameter, the weight of each line in the sum can finally be any relative integer.
So, the existence of a flux relating a set of parameters is equivalent to the existence of a vanishing linear combination of their lines.

We can now give the proof of Eq.~\ref{eq:nb_fluxes} by induction.
The relation Eq.~\ref{eq:nb_fluxes} is true for one parameter:
\begin{equation}
 M=\begin{pmatrix}
   1&1\\
   1&1
  \end{pmatrix}
 \quad\rm{or}\quad M=\begin{pmatrix}
   1&-1\\
   -1&1
  \end{pmatrix}
\end{equation}
We suppose now Eq.~\ref{eq:nb_fluxes} true for $n_\mathcal A+n_\mathcal B$ parameter and we add a parameter on a bond (possibly with a new site).
\begin{itemize}
 \item If a new site is added, the matrix gains a column and a line with a $1$ at their intersection, so the rank of $M$ increases by $1$ and $f$ remains the same.
 As we can chose the gauge on the new site, the new parameter can be chosen real, and Eq.~\ref{eq:nb_fluxes} remains true.
 \item If no new site is added and there is new flux using the new parameter, the new line is a linear combination of previous lines, thus, $\rm{rank}(M)$ is unchanged and $f$ increases by 1.
 \item If no new site is added and no new flux exists using the new parameter, the new line can not be written as a linear combination of previous lines and $f$ remains the same.
\end{itemize}

\section{Example of non local fluxes breaking the lattice symmetries}
\label{app:fluxes}
\begin{figure}
\begin{center}
 \includegraphics[height=.27\textwidth]{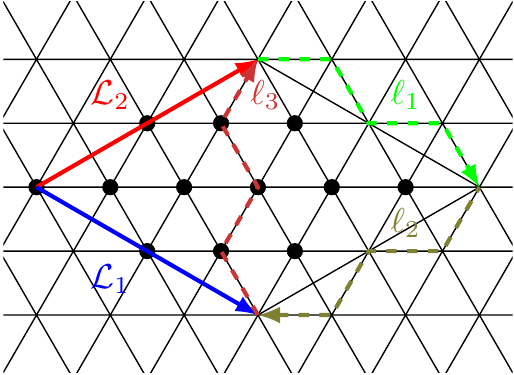}
 \caption{(Color online) Triangular lattice with 12 sites (filled circles) and periodicity $\mathcal L_1$ and $\mathcal L_2$, respecting all the lattice symmetries of Fig.~\ref{fig:sym_latt}.
  Three non local loops $\ell_1$, $\ell_2$ and $\ell_3$ are drawn in dashed arrows.
 The green dotted lines cross the bonds where the MF parameters are multiplied by $-1$.
 }
 \label{fig:non_local_loops}
\end{center}
\end{figure}

We illustrate the possibility for an Ansatz to be incompatible with a periodic lattice.
The example we give is an Ansatz on a 12-site periodic triangular lattice (Fig.~\ref{fig:non_local_loops}) that strictly respects the infinite lattice symmetries (Fig.~\ref{fig:sym_latt}).
Let us choose for simplicity the Ansatz with only first neighbor $\mathcal A_{ij}$ MF parameters defined by $k=0$ and $p_1=1$ (already discussed in the classical limit in App.~\ref{app:strange}).
This Ansatz  is the simplest illustration of these symmetry issues, but they can be encountered for any other Ansatz (as will become clearer later on).

Periodic boundary conditions defining a finite lattice are defined by the two vectors: $\mathcal L_1$ and $\mathcal L_2$. Two sites
separated by an integer linear combination of the $\mathcal L_1$ and $\mathcal L_2$ vectors (Fig.~\ref{fig:non_local_loops}) are identified as the same sites.
The three loops $\ell_1$, $\ell_2$ and $\ell_3$ are mapped onto each other by rotations and should therefore have the same fluxes in a WS or SS Ansatz.
But here their values are $0$ for $\ell_2$ and $\ell_3$ and $\pi$ for $\ell_1$.
This  is due to the fact that the unit cell of the Ansatz is twice the unit cell of the triangular lattice, and it introduces a distinction between the directions $\mathcal L_1$ and $\mathcal L_2$.
The $\cal R_\textrm{3}$ symmetry cannot be restored simply using a gauge transformation.
We see in the figure that the combination of the three loops ($\ell_1+\ell_2+\ell_3$) is a local loop, with trivial winding numbers. Thus the flux of this loop is fixed by the Ansatz and it is $\pi$. If we do not change the local physical properties of the Ansatz, which we did not want to, the sum of the three fluxes should remain equal to $\pi$ (modulo $2\pi$), and the only way out is to have a $\pi$ flux on the three non local loops.
This can be done by choosing a specific non local contour (here the green $\ell_1$ contour for example) and adding an extra phase $\pi$ to all MF bond parameters crossing this
contour.\footnote{If we had chosen a local loop the transformation would be simply be a gauge transformation with $\theta_i=\pi$ for all sites inside (or outside) the loop and
no flux (local or non-local) would be modified.}.
Transformation does not affect the local (contractible loops) fluxes since they always contain an even number of altered parameters.
But the fluxes associated to $\ell_2$ and $\ell_3$ acquire an extra phase factor.
The three fluxes around $\ell_1$, $\ell_2$ and $\ell_3$ are all equal to $\pi$ and the symmetries of the infinite lattice are now respected for this finite periodic lattice.

\begin{figure}
\begin{center}
 \includegraphics[height=.22\textwidth]{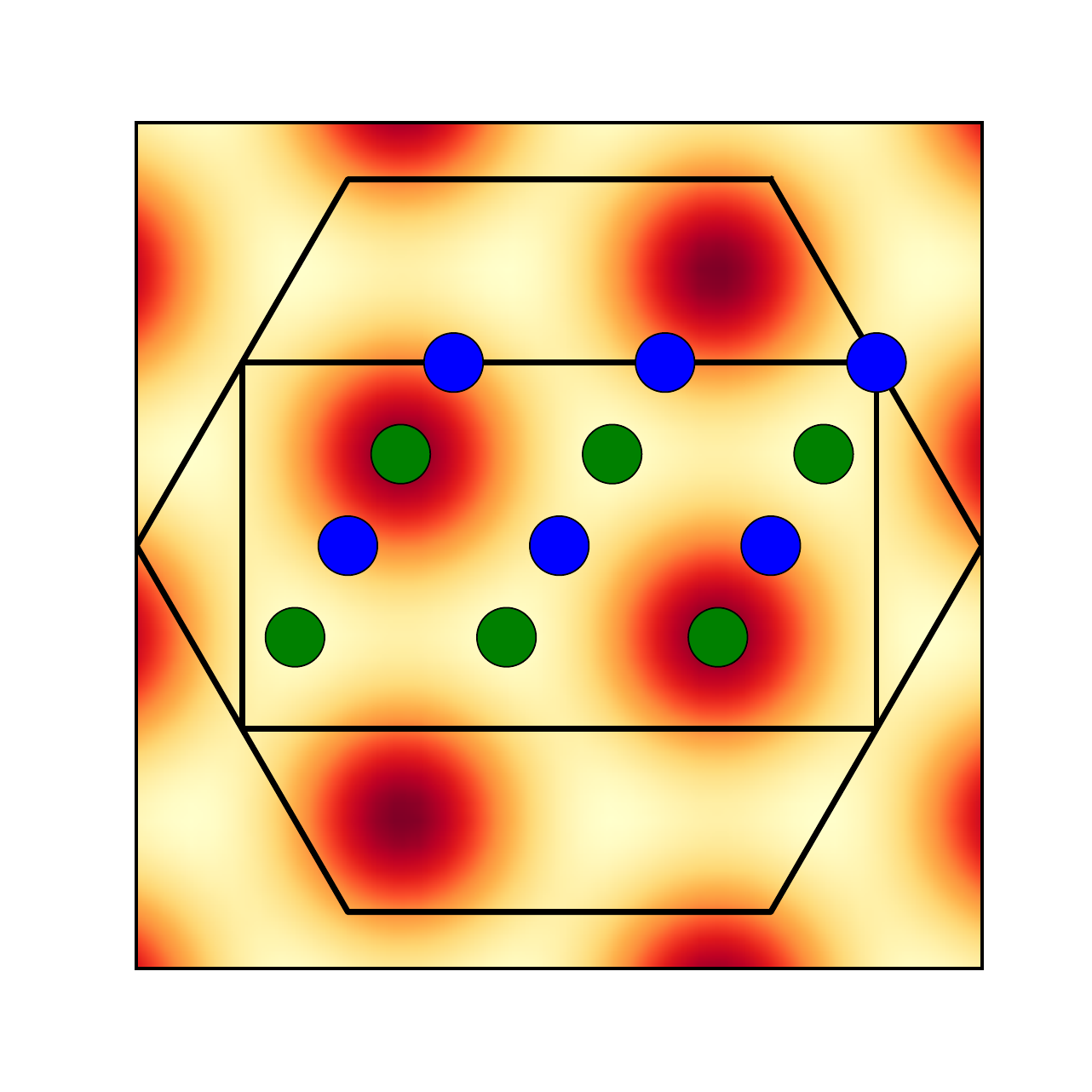}
 \caption{(Color online) The hexagon is the Brillouin zone of the triangular lattice, the rectangle, this of the Ansatz.
 The wave vectors of the 12 site lattice with periodic boundary conditions (BC's) are the blue ones, whereas they are the green ones for periodic BC in the $\mathcal L_1$ direction and antiperiodic in the $\mathcal L_2$ direction. The background intensity is the value of the spinon energy (dark for minima).
 }
 \label{fig:non_local_loops_BZ}
\end{center}
\end{figure}

We can see this modification as a change in the boundary conditions (BC's),
from periodic in both directions
$$b_{(i+n_1\mathcal L_1+n_2\mathcal L_2)\sigma}=b_{i\sigma}$$
to periodic in the $\mathcal L_1$ direction and antiperiodic in the $\mathcal L_2$ direction:
$$ \to b_{(i+n_1\mathcal L_1+n_2\mathcal L_2)\sigma}=(-1)^{n_2}b_{i\sigma},
$$
where $n_1$ and $n_2$ are arbitrary integers.
This changes the set of allowed wave vectors $\mathbf k$ from
$$
\left\{
\begin{array}{c}
 \mathbf k\cdot\mathcal L_1=0\\
 \mathbf k\cdot\mathcal L_2=0
\end{array}\right.
$$ to $$
\left\{
\begin{array}{c}
 \mathbf k\cdot\mathcal L_1=0\\
 \mathbf k\cdot\mathcal L_2=\pi
\end{array}\right..
$$
The wave vectors of the 12-site lattice before and after the transformation are drawn in Fig.~\ref{fig:non_local_loops_BZ}.
The spinon dispersion computed in the thermodynamic limit has two minima (dark red).
Periodic boundary conditions for this 12-site sample present evident drawbacks: the pattern of allowed wave vectors
(blue points in Fig.~\ref{fig:non_local_loops_BZ}) does not respect the $\cal R_\textrm{3}$ symmetry of the spinon dispersion and the minimum
of the spinon dispersion is not reached in the 12 sites samples with this PBC. We could hastily have supposed that single-spinon states are not physical
excitations and as such they do not have to respect the lattice symmetries. But this statement is incorrect. The vacuum of spinons calculated from the
set of wave vectors obtained from periodic boundary conditions  is itself distorted and so are any physical quantities as for example
spin-spin correlations that are calculated from this input.
On the contrary the modified BC's restore the ${\cal R}_6$ symmetry of the pattern of authorized wave vectors around the spinon minima.

This can also be understood in a different way. The periodic or antiperiodic boundary conditions define the 4 topological sectors on the torus.
To go from one sector to an other, we create two visons,\cite{Visons_Misguich} move one of them around the lattice and annihilate them again.
It is equivalent to the sign change of the MF parameters along this loop. The present discussion shows that for the 12-sites sample PBC do not define the
(0,0) topological sector of the model and we have to go to the APBC to describe this fully symmetric sector.
In classical terms a change of $\pi$ in a flux around a loop corresponds to a rotation of $2\pi$ of the spin orientations,
thus to a $\mathbb Z_2$ vortex.\cite{KawamuraMiyashita,KawamuraKikuchi} Translated to the classical limit, the previously used \textbf{periodic BC's}
thus correspond to a twist of $2\pi$ around the lattice, that is at the existence of a non trivial vorticity.

\bibliographystyle{apsrev4-1}
\bibliography{article_psg_chiraux}

\begin{thebibliography}{35}%
\makeatletter
\providecommand \@ifxundefined [1]{%
 \@ifx{#1\undefined}
}%
\providecommand \@ifnum [1]{%
 \ifnum #1\expandafter \@firstoftwo
 \else \expandafter \@secondoftwo
 \fi
}%
\providecommand \@ifx [1]{%
 \ifx #1\expandafter \@firstoftwo
 \else \expandafter \@secondoftwo
 \fi
}%
\providecommand \natexlab [1]{#1}%
\providecommand \enquote  [1]{``#1''}%
\providecommand \bibnamefont  [1]{#1}%
\providecommand \bibfnamefont [1]{#1}%
\providecommand \citenamefont [1]{#1}%
\providecommand \href@noop [0]{\@secondoftwo}%
\providecommand \href [0]{\begingroup \@sanitize@url \@href}%
\providecommand \@href[1]{\@@startlink{#1}\@@href}%
\providecommand \@@href[1]{\endgroup#1\@@endlink}%
\providecommand \@sanitize@url [0]{\catcode `\\12\catcode `\$12\catcode
  `\&12\catcode `\#12\catcode `\^12\catcode `\_12\catcode `\%12\relax}%
\providecommand \@@startlink[1]{}%
\providecommand \@@endlink[0]{}%
\providecommand \url  [0]{\begingroup\@sanitize@url \@url }%
\providecommand \@url [1]{\endgroup\@href {#1}{\urlprefix }}%
\providecommand \urlprefix  [0]{URL }%
\providecommand \Eprint [0]{\href }%
\providecommand \doibase [0]{http://dx.doi.org/}%
\providecommand \selectlanguage [0]{\@gobble}%
\providecommand \bibinfo  [0]{\@secondoftwo}%
\providecommand \bibfield  [0]{\@secondoftwo}%
\providecommand \translation [1]{[#1]}%
\providecommand \BibitemOpen [0]{}%
\providecommand \bibitemStop [0]{}%
\providecommand \bibitemNoStop [0]{.\EOS\space}%
\providecommand \EOS [0]{\spacefactor3000\relax}%
\providecommand \BibitemShut  [1]{\csname bibitem#1\endcsname}%
\let\auto@bib@innerbib\@empty
\bibitem [{\citenamefont {Read}\ and\ \citenamefont
  {Chakraborty}(1989)}]{Read1989a}%
  \BibitemOpen
  \bibfield  {author} {\bibinfo {author} {\bibfnamefont {N.}~\bibnamefont
  {Read}}\ and\ \bibinfo {author} {\bibfnamefont {B.}~\bibnamefont
  {Chakraborty}},\ }\href {\doibase 10.1103/PhysRevB.40.7133} {\bibfield
  {journal} {\bibinfo  {journal} {Phys. Rev. B}\ }\textbf {\bibinfo {volume}
  {40}},\ \bibinfo {pages} {7133} (\bibinfo {year} {1989})}\BibitemShut
  {NoStop}%
\bibitem [{\citenamefont {Kivelson}\ and\ \citenamefont
  {Rokhsar}(1990)}]{Kivelson1990}%
  \BibitemOpen
  \bibfield  {author} {\bibinfo {author} {\bibfnamefont {S.~A.}\ \bibnamefont
  {Kivelson}}\ and\ \bibinfo {author} {\bibfnamefont {D.~S.}\ \bibnamefont
  {Rokhsar}},\ }\href {\doibase 10.1103/PhysRevB.41.11693} {\bibfield
  {journal} {\bibinfo  {journal} {Phys. Rev. B}\ }\textbf {\bibinfo {volume}
  {41}},\ \bibinfo {pages} {11693} (\bibinfo {year} {1990})}\BibitemShut
  {NoStop}%
\bibitem [{\citenamefont {Wen}(2002)}]{Wen_PSG}%
  \BibitemOpen
  \bibfield  {author} {\bibinfo {author} {\bibfnamefont {X.-G.}\ \bibnamefont
  {Wen}},\ }\href {\doibase 10.1103/PhysRevB.65.165113} {\bibfield  {journal}
  {\bibinfo  {journal} {Phys. Rev. B}\ }\textbf {\bibinfo {volume} {65}},\
  \bibinfo {pages} {165113} (\bibinfo {year} {2002})}\BibitemShut {NoStop}%
\bibitem [{\citenamefont {Wang}\ and\ \citenamefont {Vishwanath}(2006)}]{PSG}%
  \BibitemOpen
  \bibfield  {author} {\bibinfo {author} {\bibfnamefont {F.}~\bibnamefont
  {Wang}}\ and\ \bibinfo {author} {\bibfnamefont {A.}~\bibnamefont
  {Vishwanath}},\ }\href {\doibase 10.1103/PhysRevB.74.174423} {\bibfield
  {journal} {\bibinfo  {journal} {Phys. Rev. B}\ }\textbf {\bibinfo {volume}
  {74}},\ \bibinfo {pages} {174423} (\bibinfo {year} {2006})}\BibitemShut
  {NoStop}%
\bibitem [{\citenamefont {Kalmeyer}\ and\ \citenamefont
  {Laughlin}(1987)}]{Kalmeyer_Laughlin}%
  \BibitemOpen
  \bibfield  {author} {\bibinfo {author} {\bibfnamefont {V.}~\bibnamefont
  {Kalmeyer}}\ and\ \bibinfo {author} {\bibfnamefont {R.~B.}\ \bibnamefont
  {Laughlin}},\ }\href {\doibase 10.1103/PhysRevLett.59.2095} {\bibfield
  {journal} {\bibinfo  {journal} {Phys. Rev. Lett.}\ }\textbf {\bibinfo
  {volume} {59}},\ \bibinfo {pages} {2095} (\bibinfo {year}
  {1987})}\BibitemShut {NoStop}%
\bibitem [{\citenamefont {Kalmeyer}\ and\ \citenamefont
  {Laughlin}(1989)}]{Kalmeyer_89}%
  \BibitemOpen
  \bibfield  {author} {\bibinfo {author} {\bibfnamefont {V.}~\bibnamefont
  {Kalmeyer}}\ and\ \bibinfo {author} {\bibfnamefont {R.~B.}\ \bibnamefont
  {Laughlin}},\ }\href {\doibase 10.1103/PhysRevB.39.11879} {\bibfield
  {journal} {\bibinfo  {journal} {Phys. Rev. B}\ }\textbf {\bibinfo {volume}
  {39}},\ \bibinfo {pages} {11879} (\bibinfo {year} {1989})}\BibitemShut
  {NoStop}%
\bibitem [{\citenamefont {Wen}\ \emph {et~al.}(1989)\citenamefont {Wen},
  \citenamefont {Wilczek},\ and\ \citenamefont {Zee}}]{Wen_Wilczek}%
  \BibitemOpen
  \bibfield  {author} {\bibinfo {author} {\bibfnamefont {X.-G.}\ \bibnamefont
  {Wen}}, \bibinfo {author} {\bibfnamefont {F.}~\bibnamefont {Wilczek}}, \ and\
  \bibinfo {author} {\bibfnamefont {A.}~\bibnamefont {Zee}},\ }\href {\doibase
  10.1103/PhysRevB.39.11413} {\bibfield  {journal} {\bibinfo  {journal} {Phys.
  Rev. B}\ }\textbf {\bibinfo {volume} {39}},\ \bibinfo {pages} {11413}
  (\bibinfo {year} {1989})}\BibitemShut {NoStop}%
\bibitem [{\citenamefont {Yang}\ \emph {et~al.}(1993)\citenamefont {Yang},
  \citenamefont {Warman},\ and\ \citenamefont {Girvin}}]{Yang1993}%
  \BibitemOpen
  \bibfield  {author} {\bibinfo {author} {\bibfnamefont {K.}~\bibnamefont
  {Yang}}, \bibinfo {author} {\bibfnamefont {L.~K.}\ \bibnamefont {Warman}}, \
  and\ \bibinfo {author} {\bibfnamefont {S.~M.}\ \bibnamefont {Girvin}},\
  }\href {\doibase 10.1103/PhysRevLett.70.2641} {\bibfield  {journal} {\bibinfo
   {journal} {Phys. Rev. Lett.}\ }\textbf {\bibinfo {volume} {70}},\ \bibinfo
  {pages} {2641} (\bibinfo {year} {1993})}\BibitemShut {NoStop}%
\bibitem [{\citenamefont {Messio}\ \emph {et~al.}(2011)\citenamefont {Messio},
  \citenamefont {Lhuillier},\ and\ \citenamefont {Misguich}}]{Regular_order}%
  \BibitemOpen
  \bibfield  {author} {\bibinfo {author} {\bibfnamefont {L.}~\bibnamefont
  {Messio}}, \bibinfo {author} {\bibfnamefont {C.}~\bibnamefont {Lhuillier}}, \
  and\ \bibinfo {author} {\bibfnamefont {G.}~\bibnamefont {Misguich}},\ }\href
  {\doibase 10.1103/PhysRevB.83.184401} {\bibfield  {journal} {\bibinfo
  {journal} {Phys. Rev. B}\ }\textbf {\bibinfo {volume} {83}},\ \bibinfo
  {pages} {184401} (\bibinfo {year} {2011})}\BibitemShut {NoStop}%
\bibitem [{\citenamefont {Kubo}\ and\ \citenamefont
  {Momoi}(1997)}]{Momoi_classique}%
  \BibitemOpen
  \bibfield  {author} {\bibinfo {author} {\bibfnamefont {K.}~\bibnamefont
  {Kubo}}\ and\ \bibinfo {author} {\bibfnamefont {T.}~\bibnamefont {Momoi}},\
  }\href {\doibase 10.1007/s002570050403} {\bibfield  {journal} {\bibinfo
  {journal} {Z. Phys. B}\ }\textbf {\bibinfo {volume} {103}},\ \bibinfo {pages}
  {485} (\bibinfo {year} {1997})}\BibitemShut {NoStop}%
\bibitem [{\citenamefont {Domenge}\ \emph {et~al.}(2005)\citenamefont
  {Domenge}, \citenamefont {Sindzingre}, \citenamefont {Lhuillier},\ and\
  \citenamefont {Pierre}}]{KagomeDomenge}%
  \BibitemOpen
  \bibfield  {author} {\bibinfo {author} {\bibfnamefont {J.-C.}\ \bibnamefont
  {Domenge}}, \bibinfo {author} {\bibfnamefont {P.}~\bibnamefont {Sindzingre}},
  \bibinfo {author} {\bibfnamefont {C.}~\bibnamefont {Lhuillier}}, \ and\
  \bibinfo {author} {\bibfnamefont {L.}~\bibnamefont {Pierre}},\ }\href
  {\doibase 10.1103/PhysRevB.72.024433} {\bibfield  {journal} {\bibinfo
  {journal} {Phys. Rev. B}\ }\textbf {\bibinfo {volume} {72}},\ \bibinfo
  {pages} {024433} (\bibinfo {year} {2005})}\BibitemShut {NoStop}%
\bibitem [{\citenamefont {Messio}\ \emph {et~al.}(2012)\citenamefont {Messio},
  \citenamefont {Bernu},\ and\ \citenamefont {Lhuillier}}]{cuboc1}%
  \BibitemOpen
  \bibfield  {author} {\bibinfo {author} {\bibfnamefont {L.}~\bibnamefont
  {Messio}}, \bibinfo {author} {\bibfnamefont {B.}~\bibnamefont {Bernu}}, \
  and\ \bibinfo {author} {\bibfnamefont {C.}~\bibnamefont {Lhuillier}},\ }\href
  {\doibase 10.1103/PhysRevLett.108.207204} {\bibfield  {journal} {\bibinfo
  {journal} {Phys. Rev. Lett.}\ }\textbf {\bibinfo {volume} {108}},\ \bibinfo
  {pages} {207204} (\bibinfo {year} {2012})}\BibitemShut {NoStop}%
\bibitem [{\citenamefont {F\aa{}k}\ \emph {et~al.}(2012)\citenamefont
  {F\aa{}k}, \citenamefont {Kermarrec}, \citenamefont {Messio}, \citenamefont
  {Bernu}, \citenamefont {Lhuillier}, \citenamefont {Bert}, \citenamefont
  {Mendels}, \citenamefont {Koteswararao}, \citenamefont {Bouquet},
  \citenamefont {Ollivier}, \citenamefont {Hillier}, \citenamefont {Amato},
  \citenamefont {Colman},\ and\ \citenamefont {Wills}}]{Faak2012}%
  \BibitemOpen
  \bibfield  {author} {\bibinfo {author} {\bibfnamefont {B.}~\bibnamefont
  {F\aa{}k}}, \bibinfo {author} {\bibfnamefont {E.}~\bibnamefont {Kermarrec}},
  \bibinfo {author} {\bibfnamefont {L.}~\bibnamefont {Messio}}, \bibinfo
  {author} {\bibfnamefont {B.}~\bibnamefont {Bernu}}, \bibinfo {author}
  {\bibfnamefont {C.}~\bibnamefont {Lhuillier}}, \bibinfo {author}
  {\bibfnamefont {F.}~\bibnamefont {Bert}}, \bibinfo {author} {\bibfnamefont
  {P.}~\bibnamefont {Mendels}}, \bibinfo {author} {\bibfnamefont
  {B.}~\bibnamefont {Koteswararao}}, \bibinfo {author} {\bibfnamefont
  {F.}~\bibnamefont {Bouquet}}, \bibinfo {author} {\bibfnamefont
  {J.}~\bibnamefont {Ollivier}}, \bibinfo {author} {\bibfnamefont {A.~D.}\
  \bibnamefont {Hillier}}, \bibinfo {author} {\bibfnamefont {A.}~\bibnamefont
  {Amato}}, \bibinfo {author} {\bibfnamefont {R.~H.}\ \bibnamefont {Colman}}, \
  and\ \bibinfo {author} {\bibfnamefont {A.~S.}\ \bibnamefont {Wills}},\ }\href
  {\doibase 10.1103/PhysRevLett.109.037208} {\bibfield  {journal} {\bibinfo
  {journal} {Phys. Rev. Lett.}\ }\textbf {\bibinfo {volume} {109}},\ \bibinfo
  {pages} {037208} (\bibinfo {year} {2012})}\BibitemShut {NoStop}%
\bibitem [{\citenamefont {Wen}\ \emph {et~al.}(2010)\citenamefont {Wen},
  \citenamefont {R\"uegg}, \citenamefont {Wang},\ and\ \citenamefont
  {Fiete}}]{wen2010}%
  \BibitemOpen
  \bibfield  {author} {\bibinfo {author} {\bibfnamefont {J.}~\bibnamefont
  {Wen}}, \bibinfo {author} {\bibfnamefont {A.}~\bibnamefont {R\"uegg}},
  \bibinfo {author} {\bibfnamefont {C.-C.~J.}\ \bibnamefont {Wang}}, \ and\
  \bibinfo {author} {\bibfnamefont {G.~A.}\ \bibnamefont {Fiete}},\ }\href
  {\doibase 10.1103/PhysRevB.82.075125} {\bibfield  {journal} {\bibinfo
  {journal} {Phys. Rev. B}\ }\textbf {\bibinfo {volume} {82}},\ \bibinfo
  {pages} {075125} (\bibinfo {year} {2010})}\BibitemShut {NoStop}%
\bibitem [{\citenamefont {Chua}\ \emph {et~al.}(2011)\citenamefont {Chua},
  \citenamefont {Yao},\ and\ \citenamefont {Fiete}}]{chu2011}%
  \BibitemOpen
  \bibfield  {author} {\bibinfo {author} {\bibfnamefont {V.}~\bibnamefont
  {Chua}}, \bibinfo {author} {\bibfnamefont {H.}~\bibnamefont {Yao}}, \ and\
  \bibinfo {author} {\bibfnamefont {G.~A.}\ \bibnamefont {Fiete}},\ }\href
  {\doibase 10.1103/PhysRevB.83.180412} {\bibfield  {journal} {\bibinfo
  {journal} {Phys. Rev. B}\ }\textbf {\bibinfo {volume} {83}},\ \bibinfo
  {pages} {180412} (\bibinfo {year} {2011})}\BibitemShut {NoStop}%
\bibitem [{\citenamefont {Auerbach}(1994)}]{Auerbach}%
  \BibitemOpen
  \bibfield  {author} {\bibinfo {author} {\bibfnamefont {A.}~\bibnamefont
  {Auerbach}},\ }\href@noop {} {\emph {\bibinfo {title} {{I}nteracting
  {E}lectrons and {Q}uantum {M}agnetism}}}\ (\bibinfo  {publisher}
  {Springer-Verlag, Berlin},\ \bibinfo {year} {1994})\BibitemShut {NoStop}%
\bibitem [{\citenamefont {Read}\ and\ \citenamefont
  {Sachdev}(1991)}]{ReadSachdev_SpN}%
  \BibitemOpen
  \bibfield  {author} {\bibinfo {author} {\bibfnamefont {N.}~\bibnamefont
  {Read}}\ and\ \bibinfo {author} {\bibfnamefont {S.}~\bibnamefont {Sachdev}},\
  }\href {\doibase 10.1103/PhysRevLett.66.1773} {\bibfield  {journal} {\bibinfo
   {journal} {Phys. Rev. Lett.}\ }\textbf {\bibinfo {volume} {66}},\ \bibinfo
  {pages} {1773} (\bibinfo {year} {1991})}\BibitemShut {NoStop}%
\bibitem [{\citenamefont {Trumper}\ \emph {et~al.}(1997)\citenamefont
  {Trumper}, \citenamefont {Manuel}, \citenamefont {Gazza},\ and\ \citenamefont
  {Ceccatto}}]{Trumper_fluct}%
  \BibitemOpen
  \bibfield  {author} {\bibinfo {author} {\bibfnamefont {A.~E.}\ \bibnamefont
  {Trumper}}, \bibinfo {author} {\bibfnamefont {L.~O.}\ \bibnamefont {Manuel}},
  \bibinfo {author} {\bibfnamefont {C.~J.}\ \bibnamefont {Gazza}}, \ and\
  \bibinfo {author} {\bibfnamefont {H.~A.}\ \bibnamefont {Ceccatto}},\ }\href
  {\doibase 10.1103/PhysRevLett.78.2216} {\bibfield  {journal} {\bibinfo
  {journal} {Phys. Rev. Lett.}\ }\textbf {\bibinfo {volume} {78}},\ \bibinfo
  {pages} {2216} (\bibinfo {year} {1997})}\BibitemShut {NoStop}%
\bibitem [{\citenamefont {Mezio}\ \emph {et~al.}(2011)\citenamefont {Mezio},
  \citenamefont {Sposetti}, \citenamefont {Manuel},\ and\ \citenamefont
  {Trumper}}]{Trumper_AB_SBMFT}%
  \BibitemOpen
  \bibfield  {author} {\bibinfo {author} {\bibfnamefont {A.}~\bibnamefont
  {Mezio}}, \bibinfo {author} {\bibfnamefont {C.~N.}\ \bibnamefont {Sposetti}},
  \bibinfo {author} {\bibfnamefont {L.~O.}\ \bibnamefont {Manuel}}, \ and\
  \bibinfo {author} {\bibfnamefont {A.~E.}\ \bibnamefont {Trumper}},\ }\href
  {\doibase 10.1209/0295-5075/94/47001} {\bibfield  {journal} {\bibinfo
  {journal} {EPL (Europhysics Letters)}\ }\textbf {\bibinfo {volume} {94}},\
  \bibinfo {pages} {47001} (\bibinfo {year} {2011})}\BibitemShut {NoStop}%
\bibitem [{\citenamefont {Ceccatto}\ \emph {et~al.}(1993)\citenamefont
  {Ceccatto}, \citenamefont {Gazza},\ and\ \citenamefont
  {Trumper}}]{CecattoSquare}%
  \BibitemOpen
  \bibfield  {author} {\bibinfo {author} {\bibfnamefont {H.~A.}\ \bibnamefont
  {Ceccatto}}, \bibinfo {author} {\bibfnamefont {C.~J.}\ \bibnamefont {Gazza}},
  \ and\ \bibinfo {author} {\bibfnamefont {A.~E.}\ \bibnamefont {Trumper}},\
  }\href {\doibase 10.1103/PhysRevB.47.12329} {\bibfield  {journal} {\bibinfo
  {journal} {Phys. Rev. B}\ }\textbf {\bibinfo {volume} {47}},\ \bibinfo
  {pages} {12329} (\bibinfo {year} {1993})}\BibitemShut {NoStop}%
\bibitem [{\citenamefont {Flint}\ and\ \citenamefont
  {Coleman}(2009)}]{Symplectic_SBMFT}%
  \BibitemOpen
  \bibfield  {author} {\bibinfo {author} {\bibfnamefont {R.}~\bibnamefont
  {Flint}}\ and\ \bibinfo {author} {\bibfnamefont {P.}~\bibnamefont
  {Coleman}},\ }\href {\doibase 10.1103/PhysRevB.79.014424} {\bibfield
  {journal} {\bibinfo  {journal} {Phys. Rev. B}\ }\textbf {\bibinfo {volume}
  {79}},\ \bibinfo {pages} {014424} (\bibinfo {year} {2009})}\BibitemShut
  {NoStop}%
\bibitem [{\citenamefont {{Misguich}}(2012)}]{SBMFT_alllinks}%
  \BibitemOpen
  \bibfield  {author} {\bibinfo {author} {\bibfnamefont {G.}~\bibnamefont
  {{Misguich}}},\ }\href {http://adsabs.harvard.edu/abs/2012arXiv1207.4058M}
  {\bibfield  {journal} {\bibinfo  {journal} {ArXiv e-prints}\ } (\bibinfo
  {year} {2012})},\ \Eprint {http://arxiv.org/abs/1207.4058} {arXiv:1207.4058
  [cond-mat.str-el]} \BibitemShut {NoStop}%
\bibitem [{\citenamefont {Sachdev}(1992)}]{Sachdev}%
  \BibitemOpen
  \bibfield  {author} {\bibinfo {author} {\bibfnamefont {S.}~\bibnamefont
  {Sachdev}},\ }\href {\doibase 10.1103/PhysRevB.45.12377} {\bibfield
  {journal} {\bibinfo  {journal} {Phys. Rev. B}\ }\textbf {\bibinfo {volume}
  {45}},\ \bibinfo {pages} {12377} (\bibinfo {year} {1992})}\BibitemShut
  {NoStop}%
\bibitem [{\citenamefont {Janson}\ \emph {et~al.}(2008)\citenamefont {Janson},
  \citenamefont {Richter},\ and\ \citenamefont {Rosner}}]{Janson_2008}%
  \BibitemOpen
  \bibfield  {author} {\bibinfo {author} {\bibfnamefont {O.}~\bibnamefont
  {Janson}}, \bibinfo {author} {\bibfnamefont {J.}~\bibnamefont {Richter}}, \
  and\ \bibinfo {author} {\bibfnamefont {H.}~\bibnamefont {Rosner}},\ }\href
  {\doibase 10.1103/PhysRevLett.101.106403} {\bibfield  {journal} {\bibinfo
  {journal} {Phys. Rev. Lett.}\ }\textbf {\bibinfo {volume} {101}},\ \bibinfo
  {pages} {106403} (\bibinfo {year} {2008})}\BibitemShut {NoStop}%
\bibitem [{\citenamefont {Chubukov}\ and\ \citenamefont
  {Jolicoeur}(1992)}]{Jolicoeur_J1J2tri}%
  \BibitemOpen
  \bibfield  {author} {\bibinfo {author} {\bibfnamefont {A.~V.}\ \bibnamefont
  {Chubukov}}\ and\ \bibinfo {author} {\bibfnamefont {T.}~\bibnamefont
  {Jolicoeur}},\ }\href {\doibase 10.1103/PhysRevB.46.11137} {\bibfield
  {journal} {\bibinfo  {journal} {Phys. Rev. B}\ }\textbf {\bibinfo {volume}
  {46}},\ \bibinfo {pages} {11137} (\bibinfo {year} {1992})}\BibitemShut
  {NoStop}%
\bibitem [{\citenamefont {Lecheminant}\ \emph {et~al.}(1995)\citenamefont
  {Lecheminant}, \citenamefont {Bernu}, \citenamefont {Lhuillier},\ and\
  \citenamefont {Pierre}}]{Lecheminant1995}%
  \BibitemOpen
  \bibfield  {author} {\bibinfo {author} {\bibfnamefont {P.}~\bibnamefont
  {Lecheminant}}, \bibinfo {author} {\bibfnamefont {B.}~\bibnamefont {Bernu}},
  \bibinfo {author} {\bibfnamefont {C.}~\bibnamefont {Lhuillier}}, \ and\
  \bibinfo {author} {\bibfnamefont {L.}~\bibnamefont {Pierre}},\ }\href
  {\doibase 10.1103/PhysRevB.52.6647} {\bibfield  {journal} {\bibinfo
  {journal} {Phys. Rev. B}\ }\textbf {\bibinfo {volume} {52}},\ \bibinfo
  {pages} {6647} (\bibinfo {year} {1995})}\BibitemShut {NoStop}%
\bibitem [{\citenamefont {Misguich}\ \emph {et~al.}(1998)\citenamefont
  {Misguich}, \citenamefont {Bernu},\ and\ \citenamefont
  {Lhuillier}}]{Misguich_SBMFT}%
  \BibitemOpen
  \bibfield  {author} {\bibinfo {author} {\bibfnamefont {G.}~\bibnamefont
  {Misguich}}, \bibinfo {author} {\bibfnamefont {B.}~\bibnamefont {Bernu}}, \
  and\ \bibinfo {author} {\bibfnamefont {C.}~\bibnamefont {Lhuillier}},\ }\href
  {\doibase 10.1023/A:1022588817636} {\bibfield  {journal} {\bibinfo  {journal}
  {Journal of Low Temperature Physics}\ }\textbf {\bibinfo {volume} {110}},\
  \bibinfo {pages} {327} (\bibinfo {year} {1998})}\BibitemShut {NoStop}%
\bibitem [{\citenamefont {Misguich}\ \emph {et~al.}(1999)\citenamefont
  {Misguich}, \citenamefont {Lhuillier}, \citenamefont {Bernu},\ and\
  \citenamefont {Waldtmann}}]{Misguich1999}%
  \BibitemOpen
  \bibfield  {author} {\bibinfo {author} {\bibfnamefont {G.}~\bibnamefont
  {Misguich}}, \bibinfo {author} {\bibfnamefont {C.}~\bibnamefont {Lhuillier}},
  \bibinfo {author} {\bibfnamefont {B.}~\bibnamefont {Bernu}}, \ and\ \bibinfo
  {author} {\bibfnamefont {C.}~\bibnamefont {Waldtmann}},\ }\href {\doibase
  10.1103/PhysRevB.60.1064} {\bibfield  {journal} {\bibinfo  {journal} {Phys.
  Rev. B}\ }\textbf {\bibinfo {volume} {60}},\ \bibinfo {pages} {1064}
  (\bibinfo {year} {1999})}\BibitemShut {NoStop}%
\bibitem [{\citenamefont {Misguich}\ \emph {et~al.}(2002)\citenamefont
  {Misguich}, \citenamefont {Lhuillier}, \citenamefont {Mambrini},\ and\
  \citenamefont {Sindzingre}}]{Misguichmambrini2002}%
  \BibitemOpen
  \bibfield  {author} {\bibinfo {author} {\bibfnamefont {G.}~\bibnamefont
  {Misguich}}, \bibinfo {author} {\bibfnamefont {C.}~\bibnamefont {Lhuillier}},
  \bibinfo {author} {\bibfnamefont {M.}~\bibnamefont {Mambrini}}, \ and\
  \bibinfo {author} {\bibfnamefont {P.}~\bibnamefont {Sindzingre}},\ }\href
  {\doibase 10.1140/epjb/e20020078} {\bibfield  {journal} {\bibinfo  {journal}
  {Eur. Phys. J. B}\ }\textbf {\bibinfo {volume} {26}},\ \bibinfo {pages} {167}
  (\bibinfo {year} {2002})}\BibitemShut {NoStop}%
\bibitem [{\citenamefont {Domenge}\ \emph {et~al.}(2008)\citenamefont
  {Domenge}, \citenamefont {Lhuillier}, \citenamefont {Messio}, \citenamefont
  {Pierre},\ and\ \citenamefont {Viot}}]{MessioDomenge}%
  \BibitemOpen
  \bibfield  {author} {\bibinfo {author} {\bibfnamefont {J.-C.}\ \bibnamefont
  {Domenge}}, \bibinfo {author} {\bibfnamefont {C.}~\bibnamefont {Lhuillier}},
  \bibinfo {author} {\bibfnamefont {L.}~\bibnamefont {Messio}}, \bibinfo
  {author} {\bibfnamefont {L.}~\bibnamefont {Pierre}}, \ and\ \bibinfo {author}
  {\bibfnamefont {P.}~\bibnamefont {Viot}},\ }\href {\doibase
  10.1103/PhysRevB.77.172413} {\bibfield  {journal} {\bibinfo  {journal} {Phys.
  Rev. B}\ }\textbf {\bibinfo {volume} {77}},\ \bibinfo {pages} {172413}
  (\bibinfo {year} {2008})}\BibitemShut {NoStop}%
\bibitem [{\citenamefont {Messio}\ \emph {et~al.}(2008)\citenamefont {Messio},
  \citenamefont {Domenge}, \citenamefont {Lhuillier}, \citenamefont {Pierre},
  \citenamefont {Viot},\ and\ \citenamefont {Misguich}}]{Triedres}%
  \BibitemOpen
  \bibfield  {author} {\bibinfo {author} {\bibfnamefont {L.}~\bibnamefont
  {Messio}}, \bibinfo {author} {\bibfnamefont {J.-C.}\ \bibnamefont {Domenge}},
  \bibinfo {author} {\bibfnamefont {C.}~\bibnamefont {Lhuillier}}, \bibinfo
  {author} {\bibfnamefont {L.}~\bibnamefont {Pierre}}, \bibinfo {author}
  {\bibfnamefont {P.}~\bibnamefont {Viot}}, \ and\ \bibinfo {author}
  {\bibfnamefont {G.}~\bibnamefont {Misguich}},\ }\href {\doibase
  10.1103/PhysRevB.78.054435} {\bibfield  {journal} {\bibinfo  {journal} {Phys.
  Rev. B}\ }\textbf {\bibinfo {volume} {78}},\ \bibinfo {pages} {054435}
  (\bibinfo {year} {2008})}\BibitemShut {NoStop}%
\bibitem [{\citenamefont {Colpa}(1978)}]{Colpa}%
  \BibitemOpen
  \bibfield  {author} {\bibinfo {author} {\bibfnamefont {J.~H.~P.}\
  \bibnamefont {Colpa}},\ }\href {\doibase 10.1016/0378-4371(78)90160-7}
  {\bibfield  {journal} {\bibinfo  {journal} {Physica A Statistical Mechanics
  and its Applications}\ }\textbf {\bibinfo {volume} {93}},\ \bibinfo {pages}
  {327} (\bibinfo {year} {1978})}\BibitemShut {NoStop}%
\bibitem [{\citenamefont {Misguich}\ and\ \citenamefont
  {Mila}(2008)}]{Visons_Misguich}%
  \BibitemOpen
  \bibfield  {author} {\bibinfo {author} {\bibfnamefont {G.}~\bibnamefont
  {Misguich}}\ and\ \bibinfo {author} {\bibfnamefont {F.}~\bibnamefont
  {Mila}},\ }\href {\doibase 10.1103/PhysRevB.77.134421} {\bibfield  {journal}
  {\bibinfo  {journal} {Phys. Rev. B}\ }\textbf {\bibinfo {volume} {77}},\
  \bibinfo {pages} {134421} (\bibinfo {year} {2008})}\BibitemShut {NoStop}%
\bibitem [{\citenamefont {Kawamura}\ and\ \citenamefont
  {Miyashita}(1984)}]{KawamuraMiyashita}%
  \BibitemOpen
  \bibfield  {author} {\bibinfo {author} {\bibfnamefont {H.}~\bibnamefont
  {Kawamura}}\ and\ \bibinfo {author} {\bibfnamefont {S.}~\bibnamefont
  {Miyashita}},\ }\href {\doibase 10.1143/JPSJ.53.4138} {\bibfield  {journal}
  {\bibinfo  {journal} {J. Phys. Soc. Jpn}\ }\textbf {\bibinfo {volume} {53}},\
  \bibinfo {pages} {4138} (\bibinfo {year} {1984})}\BibitemShut {NoStop}%
\bibitem [{\citenamefont {Kawamura}\ and\ \citenamefont
  {Kikuchi}(1993)}]{KawamuraKikuchi}%
  \BibitemOpen
  \bibfield  {author} {\bibinfo {author} {\bibfnamefont {H.}~\bibnamefont
  {Kawamura}}\ and\ \bibinfo {author} {\bibfnamefont {M.}~\bibnamefont
  {Kikuchi}},\ }\href {\doibase 10.1103/PhysRevB.47.1134} {\bibfield  {journal}
  {\bibinfo  {journal} {Phys. Rev. B}\ }\textbf {\bibinfo {volume} {47}},\
  \bibinfo {pages} {1134} (\bibinfo {year} {1993})}\BibitemShut {NoStop}%
\end{thebibliography}%

\end{document}